\documentclass[aps,pre,twocolumn,eqsecnum,groupedaddress]{revtex4-1}

\usepackage{amsmath,amssymb}
\usepackage{bm}
\usepackage{hyperref}
\usepackage{epsfig,graphicx}

\DeclareMathOperator{\sgn}{sgn}
\DeclareMathOperator{\sech}{sech}

\newcommand{\be}{\begin{equation}}
\newcommand{\ee}{\end{equation}}
\newcommand{\calL}{\mathcal{L}}
\newcommand{\calS}{\mathcal{S}}
\newcommand{\calH}{\mathcal{H}}

\begin{document}

\title{Typical and rare fluctuations in nonlinear driven diffusive systems with dissipation}

\author{Pablo I.\ Hurtado}
\email[]{phurtado@onsager.ugr.es}

\affiliation{Instituto Carlos I de F\'{\i}sica Te\'orica y Computacional, and Departamento de Electromagnetismo y F\'{\i}sica de la Materia, Universidad de Granada, Granada 18071, Spain}

\author{A.\ Lasanta}
\email[]{alasanta@us.es}
\affiliation{F\'{\i}sica Te\'orica, Universidad de Sevilla, Apdo.\ de Correos 1065, Sevilla 41080, Spain}

\author{A. Prados}
\email[]{prados@us.es}
\affiliation{F\'{\i}sica Te\'orica, Universidad de Sevilla, Apdo.\ de Correos 1065, Sevilla 41080, Spain}

\date{\today}

\begin{abstract}
We consider fluctuations of the dissipated energy in nonlinear driven diffusive systems subject to bulk dissipation and boundary driving. With this aim, we extend the recently-introduced macroscopic fluctuation theory to nonlinear driven dissipative media, starting from the fluctuating hydrodynamic equations describing the system mesoscopic evolution. Interestingly, the action associated to a path in mesoscopic phase-space, from which large-deviation functions for macroscopic observables can be derived, has  the same simple form as in non-dissipative systems. This is a consequence of the quasi-elasticity of microscopic dynamics, required in order to have a nontrivial competition between diffusion and dissipation at the mesoscale. Euler-Lagrange equations for the optimal density and current fields that sustain an arbitrary dissipation fluctuation are also derived.  A perturbative solution thereof shows that the probability distribution of small fluctuations is always gaussian, as expected from the central limit theorem. On the other hand, strong separation from the gaussian behavior is observed for large fluctuations, with a distribution which shows no negative branch, thus violating the Gallavotti-Cohen fluctuation theorem as expected from the irreversibility of the dynamics. The dissipation large-deviation function exhibits simple and general scaling forms for weakly and strongly dissipative systems, with large fluctuations favored in the former case but heavily supressed in the latter.  We apply our results to a general class of diffusive lattice models for which dissipation, nonlinear diffusion and driving are the key ingredients. The theoretical predictions are compared to extensive numerical simulations of the microscopic models, and excellent agreement is found. Interestingly, the large-deviation function is in some cases non-convex beyond some dissipation.
These results show that a suitable generalization of macroscopic fluctuation theory is capable of describing in detail the fluctuating behavior of nonlinear driven dissipative media.
\end{abstract}


\maketitle

\section{Introduction}
\label{s0}

Fluctuations are inherent to many physical phenomena, reflecting the hectic microscopic dynamics at macroscopic scales. In spite of their apparent random origin, essential physical information is encoded therein \cite{Landau}. A classical example is the fluctuation-dissipation theorem, which relates the linear response of a system to an external perturbation to the fluctuation properties of the system in thermal equilibrium \cite{Callen,Kubo}. More recently, the investigation of general properties of fluctuations in nonequilibrium steady states is opening new paths for understanding physics far from equilibrium \cite{Ellis,GC,LS,Bertini,Derrida,BD,Touchette,Pablo,Pablo2,Pablo3,iso}. The study of fluctuation statistics of macroscopic observables provides an alternative path to obtain thermodynamic potentials, a complementary approach to the usual ensemble description. This observation, valid both in equilibrium \cite{Landau} and nonequilibrium \cite{Bertini,Derrida}, is most relevant in the latter case because no general bottom-up approach, connecting microscopic  dynamics to macroscopic nonequilibrium properties, has been found yet. In this way, the large deviation function (LDF) controlling the statistics of these fluctuations may play in nonequilibrium statistical mechanics a role similar to the equilibrium free energy \cite{Ellis,Touchette}. A central point for this emerging paradigm is the identification of the relevant macroscopic observables characterizing the out of equilibrium behavior of the system at hand. The system dynamics often conserves locally some magnitude (a density of particles, energy, momentum, charge, etc.), and the essential nonequilibrium observable is thus the current or flux sustained by the system when subject to boundary-induced gradients or external fields. Therefore, the understanding of current statistics in terms of the microscopic dynamics represents one of the main problems of nonequilibrium statistical mechanics, triggering an intense research effort in recent years. In this context, key results are the Gallavotti-Cohen fluctuation theorem \cite{GC}, which relates the probability of observing a given current fluctuation $\vec{J}$ with the probability of the reversed event $-\vec{J}$, or the recently-introduced Isometric Fluctuation Relation \cite{iso}, which relates the probability of any pair of isometric current fluctuations $(\vec{J},\vec{J}')$, with $|\vec{J}|=|\vec{J}'|$. These intriguing ``symmetries'' appear as a consequence of the invariance under time reversal of the underlying microscopic dynamics. These and other recent results \cite{Bertini,Derrida,BD,Touchette,Pablo,Pablo2,Pablo3,GC,LS,iso} are however restricted to nonequilibrium ``conservative'' systems characterized by currents.

On the other hand, many nonequilibrium systems are inherently \textit{dissipative}, that is, they need a continuous input of energy in order to reach a steady state. In this class of systems, the relevant macroscopic observable is not only the current: the dissipated energy is also expected to play a main role. These systems include granular media \cite{BGMyR05}, dissipative biophysical systems \cite{bio}, turbulent fluids \cite{BHyP98}, active matter \cite{active}, chemical reactions \cite{chemical}, population dynamics \cite{population}, etc. In general, this class comprises all sort of reaction-diffusion systems where dissipation, diffusion and driving are the main physical mechanisms. Fluctuations in dissipative media have been much less investigated, most probably because their physics is more complicated as a result of the irreversibility of their microscopic dynamics. In principle, most of the results for nonequilibrium steady states we have referred to in the previous paragraph are not applicable to dissipative systems, since they stem from the reversibility of the underlying microscopic dynamics in the conservative case. Therefore, a question naturally arises as to whether it is possible to extend some of these ideas to dissipative media. One of the main goals of the present paper is to give a (partial) answer to this question.

In this work, we analyze both typical and rare fluctuations in nonlinear driven diffusive systems with dissipation. This is done by combining a suitable generalization of macroscopic fluctuation theory (MFT) \cite{Bertini} to the realm of dissipative media, and extensive numerical simulations of a particular albeit broad class of microscopic models. Our starting point is a general fluctuating balance equation for the (energy) density, with a drift term proportional to the spatial derivative of the current and a sink term. This mesoscopic description is expected to be valid for many driven dissipative media over a certain ``hydrodynamic'' time scale, much larger than the one characteristic of the microscopic dynamics. Over the fast (microscopic) time scale, the system forgets the initial conditions and relaxes to a local equilibrium state in which all the properties of the system become functionals of a few  ``hydrodynamic'' fields, here the density, the current and the dissipation. Afterwards, over the much slower hydrodynamic time scale, the system eventually approaches the steady state following the mesoscopic balance equation. We focus on the fluctuations of the system in this nonequilibrium steady state, in which the dissipation and the injection of energy balance each other. By using this fluctuating hydrodynamic picture together with a path integral formulation, we derive a general form for the action associated to a \emph{history} of the density, current and dissipation fields (that is, a path in mesoscopic phase space). Remarkably, this action takes the same form as in conservative nonequilibrium systems \cite{Bertini,Derrida,BD,Touchette,Pablo,Pablo2,Pablo3,iso}, simplifying the analysis in the dissipative case. This is both an important and a surprising result, which stems from the quasi-elastic character of the underlying microscopic dynamics in the large system size limit. This quasi-elasticity is necessary in order to have a balanced competition between diffusion and dissipation at the mesoscopic level. From the derived action functional, and using the recently-introduced additivity conjecture \cite{BD,Pablo,Pablo2,Pablo3}, a general form for the LDF of the dissipated energy is derived, with a  ``Lagrangian'' including second order derivatives. Therefrom, we derive the Euler-Lagrange equation (a fourth-order differential equation) for the optimal fields responsible of an arbitrary fluctuation. This Lagrangian variational problem can be mapped onto an equivalent Hamiltonian problem (four coupled first-order differential equations) which turns out to simplify the analysis. We use this Hamiltonian picture to analyze in detail three different limits, namely small fluctuations around the average for arbitrary dissipation coefficient, and the whole spectrum of fluctuations (typical and rare) for weakly- and strongly-dissipative systems. The statistics of typical (that is, small) fluctuations is gaussian as expected from the central limit theorem. However, strong separation from gaussian behavior is observed for rare fluctuations, with a distribution which shows no negative branch, thus violating the Gallavotti-Cohen fluctuation theorem as otherwise expected from the irreversible character of microscopic dynamics. We study in general the weakly-dissipative system limit using a singular perturbation expansion. This yields a simple scaling form for the dissipation LDF, showing that large dissipation fluctuations are favored in this weakly-dissipative limit, with a LDF which extends over a broad regime and decays slowly in the far positive tail. On the other hand, a different perturbative analysis in the strongly-dissipative system limit can be carried based on the formation of boundary energy layers in this limit for all fluctuations, which effectively decouples the system in two almost-independent parts. This analysis shows that large dissipation fluctuations are heavily suppressed in this limit, as opposed to the weakly-dissipative regime result.

We apply this theoretical scheme to a general class of $d$-dimensional dissipative lattice models with stochastic microscopic dynamics, for which the hydrodynamic fluctuating picture used above as starting point can be demonstrated in the large system size limit \cite{PLyH12a} (we will focus here in one dimension for simplicity). In these models there is one particle at each lattice site, characterized by its energy. Dynamics is stochastic and proceeds via collisions between nearest neighbors, at a rate which depends on the energy of the colliding pair. In a collision, a certain fraction of the pair energy is dissipated, and the remaining energy is randomly distributed within the pair. This mechanism gives rise to a nonlinear competition between diffusion and dissipation in the macroscopic limit, provided that the microscopic dissipation coefficient scales adequately with the system size. This class of models represents at a coarse-grained level the physics of many reaction-diffusion systems of technological as well as theoretical interest. In particular, when the colliding pair is chosen completely at random, independently of the value of its energy, the Kipnis-Marchioro-Presutti (KMP) model \cite{kmp} for heat conduction is recovered in the conservative case. The KMP model plays a main role in nonequilibrium statistical physics as a touchstone to test theoretical advances \cite{kmp,BD,Pablo,Pablo2,Pablo3,GC,LS,iso,KyM12}. Our general class of models contains the essential ingredients characterizing most dissipative media, namely: (i) diffusive dynamics, (ii) bulk dissipation, and (iii) boundary injection. The chances are that our results remain valid for more complex dissipative media described at the mesoscopic level by a similar evolution equation. Here we report analytical and simulation results for the statistics of the dissipated energy in this general class of models using both standard simulations and an advanced Monte Carlo method \cite{sim}. The latter allows the sampling of the tails of the distribution, and implies simulating a large number of \textit{clones} of the system.

The plan of the paper is as follows. Section \ref{s1} describes a suitable generalization of macroscopic fluctuation theory to nonlinear driven dissipative media. The large-deviation statistics of the dissipated energy, a central observable in this type of systems, is investigated here within both the Lagrangian and Hamiltonian equivalent frameworks. Section \ref{s1.5} is devoted to the detailed study of different asymptotic behaviors within the Hamiltonian formulation, which turns out to simplify the analysis. In section \ref{s2} we define a general class of microscopic lattice models whose stochastic dynamics is dissipative \cite{PLyH11a,PLyH12a}. The theoretical framework developed in the previous sections is applied to this family of models in section \ref{s4}, and the LDF for the dissipated energy is explicitly worked out. The analytical predictions are compared to extensive numerical simulations of the microscopic models, and a very good agreement is found. A summary of the main results of the paper, together with a physical discussion thereof, is given in sec. \ref{s5}. Finally, the appendix deals with some technical details that, for the sake of clarity, we have preferred to omit in the main text.

\section{Macroscopic Fluctuation Theory for Driven Dissipative Systems}
\label{s1}

In this work, we will analyze a general class of systems whose dynamics at the mesoscale is described by the following fluctuating evolution equation
\begin{equation}\label{1.1}
    \partial_t \rho(x,t)=-\partial_x j(x,t)+d(x,t) \, .
\end{equation}
We focus here in one dimension for simplicity, but our analysis can be carried out in an equivalent manner in $d$-dimensions. In eq. (\ref{1.1}), $\rho(x,t)$, $j(x,t)$ and $d(x,t)$ are the density, current and dissipation fields, respectively, and $t$ and $x\in[-1/2,1/2]$ are the macroscopic time and space variables, obtained after a diffusive scaling limit such that $x=\tilde{x}/L$ and $t=\tilde{t}/L^2$, with $\tilde{x}$ and $\tilde{t}$ the microscopic space and time variables and $L$ the system length. These coarse-grained spatial and temporal scales emerge from a suitable continuum limit of the underlying microscopic dynamics \cite{PLyH12a}. The current field is a fluctuating quantity, and can be written as
\begin{equation}\label{1.2}
    j(x,t)=-D(\rho)\partial_x \rho(x,t) +\xi(x,t).
\end{equation}
The first term is Fourier's law, where $D(\rho)$ is the diffusivity (which might be a nonlinear function of the local density), and $\xi(x,t)$ is the current noise that is gaussian and white,
\begin{equation}\label{1.3}
    \langle \xi(x,t)\rangle=0, \qquad \langle \xi(x,t) \xi(x',t')\rangle = \frac{\sigma(\rho)}{L} \delta(x-x') \delta(t-t'),
\end{equation}
with $\sigma(\rho)$ being the so-called mobility. This gaussian fluctuating field is expected to emerge for most situations in the appropriate mesoscopic limit as a result of a central limit theorem: although microscopic interactions for a given model can be highly complicated, the ensuing fluctuations of the slow hydrodynamic fields result from the sum of an enormous amount of random events at the microscale which give rise to gaussian statistics, with an amplitude of the order of $L^{-1/2}$, in the mesoscopic regime in which eq. (\ref{1.1}) emerges. On the other hand, the dissipation field $d(x,t)$ is
\begin{equation}\label{1.3b}
    d(x,t)=-\nu R(\rho(x,t)),
\end{equation}
where $\nu$ is the macroscopic dissipation coefficient, and $R(\rho)$ is a certain function of the density $\rho$. For the calculations which follow throughout this section, it is useful to introduce a new variable $y$, such that
\begin{subequations}\label{1.4}
\begin{equation}\label{1.4a}
    y=R(\rho),
\end{equation}
\begin{equation}\label{1.4b}
    d(x,t)=-\nu y(x,t).
\end{equation}
\end{subequations}
The dissipation field is present at the mesoscopic level because the microscopic stochastic dynamics of the models of interest dissipates some energy, that is, we have the equivalent of a microscopic restitution coefficient $\alpha$, so that the amount of dissipated energy is proportional to $1-\alpha$.  The macroscopic dissipation coefficient $\nu$ is thus proportional to $1-\alpha$. Note however that there is no noise term in eq. (\ref{1.3b}), so the local fluctuations of the dissipation field are enslaved to those of the density $\rho(x,t)$. The physical reason for this behavior is that the microscopic dynamics must be quasi-elastic in order to  ensure that dissipation and  diffusion take place over the same time scale in the thermodynamic limit. Typically, $1-\alpha$ must scale as $L^{-2}$ that is the order of magnitude of the diffusive term in a system of length $L$ \cite{PLyH11a,PLyH12a}.

The boundary conditions for eq (\ref{1.1}) depend on the physical situation of interest. For instance, we may consider that the system is kept in contact with two thermal reservoirs at $x=\pm 1/2$, at the same temperature $T$, so $\rho(\pm 1/2,t)=T$.  In that case, the system eventually reaches a steady state in the long time limit, for which the injection of energy through the boundaries and the dissipation balance each other. The stationary average (macroscopic) solution of (\ref{1.1}) verifies
\begin{equation}\label{1.4c}
    j_{\text{av}}'(x)+\nu R(\rho_{\text{av}}(x))=0, \quad j_{\text{av}}(x)=-D(\rho_{\text{av}}(x)) \rho_{\text{av}}'(x),
\end{equation}
where the prime indicates spatial derivative. The first equation in (\ref{1.4c}) follows from (\ref{1.1}), and the second one is Fourier's law for the averages. Equivalently, a closed second-order equation for $\rho$ may be written,
\begin{equation}\label{1.4d}
    \frac{d}{dx} \left[ D(\rho_{\text{av}}) \rho'_{\text{av}} \right]=\nu R(\rho_{\text{av}}),
\end{equation}
with the boundary conditions $\rho_{\text{av}}(\pm 1/2)=T$. Equations (\ref{1.4c}) and (\ref{1.4d}) can be also written for the variable $y$ introduced in eq. (\ref{1.4a}),
\begin{equation}\label{1.4e}
    j'_{\text{av}}(x)+\nu y_{\text{av}}(x)=0, \quad j_{\text{av}}(x)=-\hat{D}(y_{\text{av}}(x)) y_{\text{av}}'(x),
\end{equation}
with
\begin{equation}\label{1.4f}
    \hat{D}(y)=\left(\frac{dy}{d\rho}\right)^{-1}D(\rho),
\end{equation}
since
\begin{equation}\label{1.4g}
    j_{\text{av}}(x,t)=-\hat{D}(y_{\text{av}})\partial_x y(x,t).
\end{equation}
Thus $\hat{D}$ is an ``effective'' diffusivity: it is the factor multiplying the spatial gradient when writing Fourier's equation in terms of the new variable $y$. Equation (\ref{1.4e}) can also be summarized in a second order differential equation for $y_{\text{av}}$,
\begin{equation}\label{1.4h}
    \left[ \hat{D}(y_{\text{av}}) y'_{\text{av}} \right]'=\nu y_{\text{av}}, \quad y_{\text{av}}(\pm 1/2)=R(T).
\end{equation}
Interestingly, it can be shown (see below) that $\hat{D}$ is constant, independent of $y$, whenever $y=R(\rho)$ depends algebraically in $\rho$, a case  we will study in detail in section \ref{s4}. This observation considerably simplifies the subsequent analysis.

The probability of observing a history $\{\rho(x,t),j(x,t)\}_0^\tau$ of duration $\tau$ for the density and current fields, starting from a given initial state, can be written now as a path integral over all the possible realizations of the current noise $\{\xi(x,t)\}_0^\tau$, weighted by its gaussian measure, and restricted to those realizations compatible with eq. (\ref{1.1}) at every point of space and time \cite{Pablo2}. This probability hence obeys a large deviation principle of the form \cite{Bertini,Derrida,BD,Ellis,Touchette,Pablo,Pablo2,Pablo3,PLyH11a}
\begin{equation}\label{1.5}
    P(\{\rho,j\}_0^\tau)\sim \exp \left(+ L\, {\cal I}_\tau[\rho,j] \right),
\end{equation}
with a rate functional \cite{Bertini,Derrida}
\begin{equation}\label{1.6}
     {\cal I}_\tau[\rho,j]=-\int_0^\tau dt \int_{-1/2}^{1/2} dx \, \frac{[j+D(\rho)\partial_x\rho]^2}{2\sigma(\rho)}
\end{equation}
with $\rho(x,t)$ and $j(x,t)$ coupled via the balance equation (\ref{1.1}), and the dissipation $d(x,t)$ given in terms of $\rho(x,t)$ by (\ref{1.3b}). Equation (\ref{1.6}) expresses the gaussian nature of the \emph{local} current fluctuations around its average (Fourier's law) behavior. The functional in (\ref{1.6}) is the same as in the conservative case (that is, with no bulk dissipation), due to the quasi-elasticity of the microscopic dynamics, which makes the current noise be the only relevant one in the hydrodynamic description, see discussion in section \ref{s2} \cite{PLyH12a}. We focus now on the fluctuations of the dissipated energy, integrated over space and time
\begin{eqnarray}\label{1.7}
    d&=&-\frac{1}{\tau} \int_0^\tau dt \int_{-1/2}^{1/2}dx \, d(x,t)  \\
      &=&\frac{\nu}{\tau} \int_0^\tau dt \int_{-1/2}^{1/2}dx \, R(\rho(x,t)) >0 \, , \nonumber
\end{eqnarray}
where we have introduced a minus sign for the sake of convenience, in order to make $d$ positive. As discussed above, this is a fundamental observable to understand the statistical physics of driven dissipative media. The probability of such a fluctuation $P_\tau(d)$ scales in the long-time limit as
\begin{equation}\label{1.8}
    P_\tau(d)\sim \exp\left[+\tau L \,G(d) \right], \quad G(d)=\frac{1}{\tau} \max_{\rho,j}I_{\tau}[\rho,j] \, .
\end{equation}
\begin{figure}
\centerline{
\includegraphics[width=8cm]{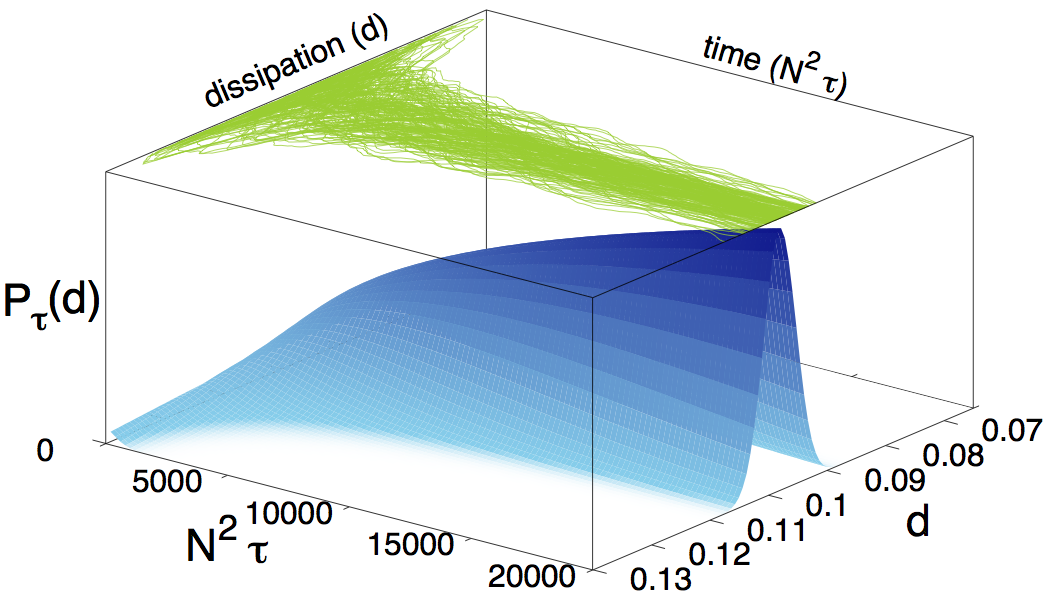}}
\caption{(Color online) Convergence of the space\&time-integrated dissipation to its ensemble value for many different realizations, and sketch of the probability concentration  as time increases, associated with the large deviation principle, eq. (\ref{1.8}).
}
\label{sketch}
\end{figure}
This defines a new large deviation principle for $d$, see Fig. \ref{sketch}, such that $G(d)$ is obtained from ${\cal I}_\tau[\rho,j]$ via a saddle-point calculation for long times (that is, it follows from the contraction of the original rate function ${\cal I}_{\tau}$ \cite{Touchette}). The optimal fields $\rho_0(x,t;d)$, $j_0(x,t;d)$ which are the solution of the variational problem (\ref{1.8}) must be consistent with the prescribed value of the dissipated energy $d$ in (\ref{1.7}), and are also related by the balance equation (\ref{1.1}), supplemented with (\ref{1.3b}) and the appropriate boundary conditions. These optimal fields can be interpreted as the ones adopted by the system to sustain a long-time fluctuation of the space\&time-integrated dissipation $d$. For the sake of simplicity, we have not explicitly introduced in our notation the parametric dependence  of the LDF $\mathbf{G(d)}$ and the associated optimal profiles on the boundary temperature $T$, though this should be borne in mind for latter reference.

\subsection{\label{s1.2a}The constrained variational problem}

We now assume that these optimal profiles do not depend on time. In conservative systems, this conjecture has been shown \cite{Bertini} to be equivalent to the additivity principle recently introduced to study current fluctuations in diffusive media \cite{BD}. The validity of this additivity scenario has been recently confirmed in extensive numerical simulations for a broad interval of fluctuations \cite{Pablo,iso}, though it may eventually break down for extreme fluctuations via a dynamic phase transition \cite{BD05,PabloSSB}. As we will see below, the applicability of this generalization of the additivity conjecture to dissipative systems is well supported by numerical evidence. Under this simplifying hypothesis, the fluctuating balance equation (\ref{1.1}) reduces to
\begin{equation}\label{1.8b}
    j'(x)+\nu y(x)=0, \quad y(x)=-j'(x)/\nu,
\end{equation}
making use of the variable $y$ defined in eq. (\ref{1.4a}). Moreover, we can integrate over time in the definition (\ref{1.7}) of the integrated dissipation $d$,
\begin{subequations}\label{1.8c}
\begin{equation}
    d=\nu \int_{-1/2}^{1/2} dx\, y(x) ,
\end{equation}
or, equivalently,
\begin{equation}
    d= -\int_{-1/2}^{1/2} dx\, j'(x)=j(-1/2)-j(1/2) >0.
\end{equation}
\end{subequations}
In this way, by using the additivity hypothesis we can eliminate $\rho(x)$ and write $G(d)$ in terms of only one variable as
\begin{subequations}\label{1.9}
\begin{equation}\label{1.9a}
    G(d)=-\min_{j(x)} {\calS}[j], \quad \text{\bf with }  \calS[j]=\int_{-1/2}^{1/2} dx \, {\cal L}(j,j'j'') \, ,
\end{equation}
\begin{equation}
    {\cal L}(j,j',j'')=\frac{[j-\hat{D}(-j'/\nu)\frac{\displaystyle j''}{\displaystyle \nu}]^2}{2\hat{\sigma}(-j'/\nu)} \, ,
\end{equation}
\end{subequations}
where $\hat{D}$ is the effective diffusivity defined in eq. (\ref{1.4f}), and $\hat{\sigma}$ is the mobility, defined in eq. (\ref{1.3}), both written in terms of $y=-j'/\nu$. The function ${\cal L}(j,j',j'')$ is a generalized Lagrangian with dependence on first and also second order derivatives, see the appendix.

We have to find the optimal current profile $j_0(x;d)$, that is, the solution of the variational problem (\ref{1.9}), with the constraint that the integrated dissipation $d$ has a definite value, as given by (\ref{1.8c}). Therefore we must use the Lagrange multiplier procedure \cite{GyF63,La49}, that is, look for an extremum of
\begin{eqnarray}
  \calS_\lambda[j] & = & \calS[j]-\lambda\int_{-1/2}^{1/2} dx \, (j'+d) \\
  & = & \int_{-1/2}^{1/2} dx \, \calL_\lambda(j,j',j'')
  \label{var1}
\end{eqnarray}
where
\begin{equation}\label{var2}
  \calL_\lambda(j,j',j'')=\calL(j,j',j'')-\lambda (j'+d),
\end{equation}
with $\lambda$ being the Lagrange multiplier. The extremum of $\calS_\lambda$ follows from two conditions: (i) $\delta \calS_\lambda=0$, and (ii) $\partial S_\lambda/\partial\lambda=0$. The first condition implies
\begin{equation}\label{var3a}
     \frac{d^2}{dx^2}\left(\frac{\partial \cal L_\lambda}{\partial j''}\right)-\frac{d}{dx}\left( \frac{\partial\cal  L_\lambda}{\partial j'}\right)+\frac{\partial \cal L_\lambda}{\partial j} =0,
\end{equation}
which is the Euler-Lagrange equation for a Lagrangian $L_\lambda$ containing second order derivatives (see the appendix). Condition (ii) leads to
the constraint on the integrated dissipation, given by Eq. (\ref{1.8c}). The boundary conditions for the Euler-Lagrange equation are
\begin{equation}\label{var4}
  j'(\pm 1/2)=-\nu R(T), \quad p_{\lambda j}(\pm 1/2)=0.
\end{equation}
We have introduced the generalized momentum $p_j$ conjugate of $j$, for the new Lagrangian $\calL_\lambda$, as
\begin{equation}\label{var5}
  p_{\lambda j}=\frac{\partial\calL_\lambda}{\partial j'}-\frac{d}{dx} \left(
  \frac{\partial\calL_\lambda}{\partial j''} \right).
\end{equation}
These boundary conditions arise from (1) the values of the density at the boundaries, which are prescribed, $\rho(\pm 1/2)=T$, and (2) the condition $\delta S_\lambda=0$, which provides the additional needed conditions when there are not enough values of the variables fixed at the boundaries (see the appendix, and also \cite{La49}).

\subsection{\label{s1.2b}Mapping the constraint to boundary conditions}

Taking into account the relation between $\calL_\lambda$ and $\calL$, Eq. (\ref{var2}), the generalized momentum $p_{\lambda j}$ verifies
\begin{equation}\label{var6}
  p_{\lambda j}=p_j-\lambda,
\end{equation}
where $p_j$ is the generalized momentum for the Lagrangian $\calL$, that is,
\begin{equation}\label{var7}
   p_{j}=\frac{\partial\calL}{\partial j'}-\frac{d}{dx} \left(
  \frac{\partial\calL}{\partial j''}\right).
\end{equation}
Moreover, the Euler-Lagrange equation (\ref{var3a}) implies that
\begin{equation}\label{var8}
   \frac{d^2}{dx^2}\left(\frac{\partial \cal L}{\partial j''}\right)-\frac{d}{dx}\left( \frac{\partial\cal  L}{\partial j'}\right)+\frac{\partial \cal L}{\partial j} =0,
\end{equation}
that is, we also obtain the Euler-Lagrange equation corresponding to the original Lagrangian $\calL$. The definition of $p_j$ in eq. (\ref{var7}) guarantees that $p'_j=\partial\calL/\partial j$, as in the case of the usual variational problem with a Lagrangian including only first-order derivatives. Now, the boundary conditions can be written as
\begin{equation}\label{var9}
  j'(\pm 1/2)=-\nu R(T), \quad p_{j}(\pm 1/2)=\lambda,
\end{equation}
which follow from Eqs. (\ref{var4}) and (\ref{var6}). The above result imply that our constrained variational problem can be mapped onto a unconstrained variational problem with the original Lagrangian $\calL$, its associated Euler-Lagrange equation (\ref{var8}) and the boundary conditions (\ref{var9}). The unknown value $\lambda$ for the generalized momentum $p_j$ at the boundaries must be determined by imposing the prescribed value of the integrated dissipation, as given by Eq. (\ref{1.8c}), that is, $\lambda=\lambda(d)$. In particular, $\lambda=0$ is equivalent to imposing no restrictions on the integrated dissipation, so that we should recover (as we will see later) the average profiles and dissipation in this case. In this sense, a non-zero value of $\lambda=p_j(\pm 1/2)$ is a measure of the departure from the average hydrodynamic behaviour.

On physical grounds, we expect the corresponding optimal density profile to be an even function of $x$, because of the symmetry of our system around the center $x=0$. In fact, the Euler-Lagrange equation (\ref{var8}) admits solutions with well-defined parity. Since the Lagrangian has the symmetry property $\calL(-j,j',-j'')=\calL(j,j',j'')$, Eq. (\ref{var8}) has solutions with $j$ being an odd function of $x$, which implies that $y$ (and therefore $\rho$) is an even function of $x$. From now on, we will restrict ourselves to these symmetric solutions of the variational problem. Thus, Eq. (\ref{1.8c}) reduces to
\begin{equation}\label{1.9b}
  d=2 j(-1/2;d)=-2j(1/2;d),
\end{equation}
so the boundary conditions for the Euler-Lagrange equation boil down to
\begin{equation}\label{1.11}
  j'(\pm 1/2;d)=-\nu R(T), \; j(-1/2;d)=-j(1/2;d)=d/2 \, ,
\end{equation}
i.e. much simpler than Eq. (\ref{var9}). This follows from $p_j$ being an even function of $x$ for the solutions with well-defined parity we are considering. In this way, symmetry considerations lead to the simpler boundary conditions (\ref{1.11}), which in turn allow us to get rid of the Lagrange multiplier $\lambda$. In summary, we have mapped our original variational problem with the subsidiary condition that the dissipation has a given value to an unconstrained variational problem, with the original Lagrangian $\calL(j,j',j'')$ and prescribed values of $j$ and $j'$ at the boundaries.

Once the optimal current profile is obtained, the optimal density profile can be calculated from the balance equation (\ref{1.8b}). Of course, the density profile so obtained obeys the boundary conditions $\rho(\pm 1/2;d)=T$. It must be stressed that the Euler-Lagrange equation (\ref{var8}) with boundary conditions (\ref{1.11}) gives the correct solution to the constrained variational problem when the optimal profiles have a well-defined parity. Nevertheless, one cannot rule out the existence of symmetry-breaking solutions without well-defined parity, since in general a variational problem may have multiple solutions \cite{GyF63}. In that case, one must solve the more complex variational problem comprising the Euler-Lagrange equation (\ref{var8}) with the boundary conditions (\ref{var9}), where the Lagrange multiplier $\lambda=\lambda(d)$ is determined by imposing the constraint (\ref{1.8c}). We note however that numerical evidence strongly supports the validity of symmetric solutions (see below).

The LDF $G(d)$ depends on $d$ and $T$ through the boundary conditions (recall that its $T$-dependence has been omitted in notation for simplicity). The derivation of the Euler-Lagrange equation (\ref{var8}) (see the appendix) shows that, for the solutions with well-defined parity,
\begin{equation}\label{var10}
  \delta G(d)=p_j(1/2) \delta d + 2 p_{j'}(1/2)\nu \frac{dR(T)}{dT} \delta T,
\end{equation}
where $p_j$ is the generalized momentum conjugate of $j$, defined in Eq. (\ref{var7}), and $p_{j'}$ is the generalized momentum conjugate of $j'$,
\begin{equation}\label{var11}
  p_{j'}=\frac{\partial\calL}{\partial j''},
\end{equation}
which is an odd function of $x$ for the solutions with well-defined parity. In this way Eq. (\ref{var10}) offers a geometric interpretation for the values of the generalized momenta at the boundaries, as they are directly related to the partial derivatives of the LDF,
\begin{equation}\label{var12}
  \frac{\partial G}{\partial d}=p_j(1/2), \quad \frac{\partial G}{\partial T}=2 \nu \frac{dR(T)}{dT} p_{j'}(1/2).
\end{equation}

\subsection{A Hamiltonian formulation of the problem}
\label{s1.3}

We will not write the detailed form of the general fourth-order differential equation (\ref{var8}) for the optimal profile $j(x;d)$, since it is not particularly illuminating. Instead we now write a set of four equivalent first order differential equations arising in the  equivalent ``Hamiltonian'' description. In the following, we sketch the procedure to introduce the Hamiltonian for a Lagrangian with higher-order derivatives \cite{La49,GyF63}, adapted to the present case. As the Euler-Lagrange equation is a fourth-order differential equation, we should have two canonical coordinates and their two corresponding canonical momenta. The first canonical coordinate is the current $j$, and we choose the second one to be $y$, which is proportional to $j'$, as given by Eq. (\ref{1.8b}). This choice is suggested by the structure of the Lagrangian in eq. (\ref{1.9}). It is worth recalling that the density profile can be directly obtained from $y$ by making use of its definition, Eq. (\ref{1.4a}). Next, we introduce the canonical momenta $p_y$ and $p_j$ conjugate to $y$ and $j$, respectively. The momentum $p_j$ has been defined in Eq. (\ref{var7}), and $p_y$ is given by
\begin{equation}\label{1.12}
    p_y\equiv-\nu\frac{\partial{\cal L}}{\partial j''},
\end{equation}
which follows from the definition of $p_{j'}$, Eq. (\ref{var11}), and Eq. (\ref{1.8b}) for $y$.
The Hamiltonian is then introduced in the usual way,
\begin{equation}\label{1.12b}
    {\cal H}\equiv y' p_y+j' p_j-{\cal L}\equiv y'p_y-\nu y p_j-{\cal L}.
\end{equation}
After some algebra, we get
\begin{subequations}\label{1.13}
\begin{equation}\label{1.13a}
    \mathcal{H}=\frac{1}{2}Q(y) p_y^2-{\hat{D}}^{-1}(y)j p_y-\nu y p_j,
\end{equation}
\begin{equation}\label{1.13b}
    Q(y)\equiv \frac{\hat{\sigma}(y)}{\hat{D}^2(y)},
\end{equation}
\end{subequations}
where we have defined the auxiliary function $Q(y)$, with $Q(y)>0$ for all $y$.  We have also made use of eq. (\ref{1.8b}), of  the expression for $p_y$ which follows from its definition (\ref{1.12}),
\begin{equation}\label{1.13c}
    p_y=\hat{D}(y)  \frac{j+\hat{D}(y)y'}{\hat{\sigma}(y)},
\end{equation}
and of the Lagrangian
\begin{equation}\label{1.13d}
    {\cal L}=\frac{\hat{\sigma}(y) p_y^2}{2 \hat{D}^2(y)}=\frac{1}{2}Q(y)p_y^2,
\end{equation}
written in terms of the canonical variables, with the aid of eq. (\ref{1.13c}). As usual, $\cal H$ is a function of $(y,j,p_y,p_j)$, which satisfy the following set of four ''canonical'' first-order differential equations,
\begin{subequations}\label{1.14}
\begin{equation}\label{1.14a}
     y'=\frac{\partial\cal H}{\partial p_y}=Q(y)p_y-{\hat{D}}^{-1}(y)j,
\end{equation}
\begin{equation}\label{1.14b}
    j'=\frac{\partial\cal H}{\partial p_j}=-\nu y,
\end{equation}
\begin{equation}\label{1.14c}
    p'_y=-\frac{\partial\cal H}{\partial y}=-\frac{dQ(y)}{dy}  \frac{p_y^2}{2}+\frac{d{\hat{D}}^{-1}(y)}{dy}j p_y+\nu p_j,
\end{equation}
\begin{equation}\label{1.14d}
    p'_j=-\frac{\partial\cal H}{\partial j}={\hat{D}}^{-1}(y)p_y
\end{equation}
\end{subequations}
that are equivalent to the fourth-order Euler-Lagrange equation (\ref{var8}). Note that, as is usual in physics and in order not to clutter our formulae, we have dropped the subindex $0$ for the optimal profiles, which are now solutions of the above canonical equations; the same notation is used for the canonical variables in the Hamiltonian and for the solutions of Hamilton's equations. On the other hand, as the Hamiltonian does not depend explicitly on $x$, it is a first integral of the system (\ref{1.14}):  $\cal H=\text{const.}$ over any of its solutions. This property  may be used to simplify the integration of the system.

In general, for a given value of the dissipation $d$, we have to solve the system of equations (\ref{1.14}) with the boundary conditions
\begin{equation}\label{1.15b}
y(\pm 1/2)=R(T), \quad j(-1/2)=-j(1/2)=d/2,
\end{equation}
which follow from Eqs. (\ref{1.11}) and (\ref{1.8b}). Again, we have to look for solutions of Eq. (\ref{1.14}) with well-defined parity, that is, $y$ and $j$ are even and odd functions of $x$, respectively (and, therefore, $p_y$ is odd and $p_j$ even). The solution of these canonical equations is then inserted into the expression of the LDF $G(d)$, which can be written in terms of the canonical variables as
\begin{equation}\label{1.15d}
    G(d)=-\int_{-1/2}^{1/2} dx \, {\cal L}=-\frac{1}{2}\int_{-1/2}^{1/2} dx \,    Q(y) p_y^2,
\end{equation}
by combining eqs. (\ref{1.9}) and (\ref{1.13d}). In this way, we obtain the LDF for an arbitrary value of the integrated dissipation $d$ within the Hamiltonian formulation of the variational problem. Equation (\ref{1.15d}) shows clearly that the most probable (average) profiles correspond to a solution with $p_y=0$ for all x, for which $G(d)$ vanishes. By substituting $p_y=0$ in Eqs. (\ref{1.14c})-(\ref{1.14d}), we also have that  $p_j=0$ for all $x$. Moreover, Eqs. (\ref{1.14a}) and (\ref{1.14b}) simplify to eq. (\ref{1.4e}), that is, the average profiles are reobtained. Therefore, there is always a solution of the canonical equations (\ref{1.14}) with identically vanishing canonical momenta, which corresponds to the average solution of the hydrodynamic equation (\ref{1.4e}) \cite{foot1}. These average hydrodynamic profiles $\{\rho_{\text{av}},j_{\text{av}}\}$ lead to the average value of the integrated dissipation
\begin{eqnarray}
    d_{\text{av}}&=&\nu \int_{-1/2}^{1/2} dx \, R(\rho_{\text{av}})=\nu \int_{-1/2}^{1/2} dx\, y_{\text{av}}(x) \nonumber \\
    & =& 2j_{\text{av}}(-1/2). \label{1.16b}
\end{eqnarray}
This discussion is consistent with the one below Eq. (\ref{var9}), which was done within the framework of the equivalent Lagrangian description. Fluctuations involve  non-zero values for the canonical momenta, whose magnitude is then a measure of the departure from the average behaviour $(d-d_{\text{av}})/d_{\text{av}}$.

\section{\label{s1.5} Analysis of the LDF in some limiting cases}

In the following subsections, we further analyze the form of the LDF in certain limits of interest, for which some general results can be obtained. First, we focus on the behavior of $G(d)$ for small fluctuations around the average, where a quadratic shape of the LDF is expected (corresponding to gaussian fluctuations). We then analyze the limit of weakly-dissipative systems, $\nu\ll 1$, for which an adequate perturbative expansion allows us to obtain a non-trivial and interesting scaling form for the LDF. Finally, we consider the opposite limit of strongly-dissipative systems, $\nu\gg 1$, for which a different scaling for the LDF is found.

\subsection{Small fluctuations around the average}
\label{s1.5a}

As the average behavior corresponds to the particular solution of the canonical equations corresponding to vanishing momenta $p_j=0$, $p_\rho=0$,
small fluctuations can be thus analyzed by assuming that the canonical momenta are small. Let us define the dimensionles parameter
\begin{equation}\label{1.16}
    \epsilon=\frac{d-d_{\text{av}}}{d_{\text{av}}}
\end{equation}
to measure the separation from the average integrated dissipation $d_{\text{av}}$.
As we have just discussed, the canonical momenta vanish for $\epsilon=0$. We write
\begin{equation}\label{1.16c}
    y=y_{\text{av}}+\epsilon \Delta y, \; j=j_{\text{av}}+\epsilon \Delta j, \; p_y=\epsilon \Delta p_y, \; p_j=\epsilon \Delta p_j,
\end{equation}
and linearize Eqs. (\ref{1.14}) around the average solution, that is, we only retain terms linear in $\epsilon$. Then,
\begin{subequations}\label{1.17}
\begin{equation}\label{1.17a}
    \Delta y'=Q(y_{\text{av}})\Delta p_y-{\hat{D}}^{-1}(y_{\text{av}}) \Delta j -j_{\text{av}} \frac{d{\hat{D}}^{-1}(y_{\text{av}})}{dy_{\text{av}}} \Delta y,
\end{equation}
\begin{equation}\label{1.17b}
    \Delta j'=-\nu \Delta y,
\end{equation}
\begin{equation}\label{1.17c}
    \Delta p'_y= j_{\text{av}} \frac{d{\hat{D}}^{-1}(y_{\text{av}})}{dy_{\text{av}}} \Delta p_y+\nu \Delta p_j,
\end{equation}
\begin{equation}\label{1.17d}
    \Delta p'_j={\hat{D}}^{-1}(y_{\text{av}}) \Delta p_y.
\end{equation}
\end{subequations}
The boundary conditions for these equations are $\Delta y(\pm 1/2)=0$, $\Delta j(-1/2)=-\Delta j(1/2)=d_{\text{av}}/2$. The solution of this system of equations must be inserted in the large deviation function (\ref{1.9}). Using the expression (\ref{1.15d}) for the LDF, it follows that
\begin{equation}\label{1.18}
    G(d)\sim-\frac{\epsilon^2}{2} \int_{-1/2}^{1/2} dx \, Q(y_{\text{av}}) \Delta p_y^2,
\end{equation}
for small fluctuations of the dissipation around the average. Taking into account (\ref{1.16}) and the large deviation principle (\ref{1.8}), eq. (\ref{1.18}) means that the probability of such small fluctuations of the integrated dissipation $d$ is approximately gaussian,
\begin{equation}\label{1.19}
    P_\tau(d) \stackrel{|\epsilon|\ll 1}{\propto} \exp \left[ -L\tau \frac{(d-d_{\text{av}})^2}{2 d_{\text{av}}^2 \Lambda_{\nu}^2}\right]
\end{equation}
with $\Lambda_{\nu}^2$ given by
\begin{equation}\label{1.19b}
    \Lambda_{\nu}^2= \left( \int_{-1/2}^{1/2} dx \,  Q(y_{\text{av}}) \Delta p_y^2\right)^{-1}.
\end{equation}
In this way, the gaussian estimation for the standard deviation of the dissipation, by comparing (\ref{1.19}) to (\ref{1.19b}), is given by $\chi\equiv d_{\text{av}}\Lambda_{\nu}/\sqrt{\tau L}$. In order to make a more detailed study of the LDF, concrete functional dependences of the diffusivity $D$, the mobility $\sigma$ and the dissipation $R$ on the density $\rho$ must be considered. This is done in the following sections of the paper, where we will consider a broad family of models for which the transport coefficients can be explicitly obtained. On the other hand, it is important to notice that gaussian statistics is only expected for small fluctuations around the average dissipation. In general, the solution of the variational problem given by the integration of eq.(\ref{1.14}), when inserted into (\ref{1.15d}), will give rise to non-gaussian statistics (that is, a non-quadratic dependence of the LDF) for an arbitrary fluctuation of the dissipated energy $d$.

\subsection{Weakly-dissipative systems, $\nu\ll 1$}
\label{s1.5b}

We proceed now by analysing the canonical equations (\ref{1.14}) in the limit $\nu\ll 1$. Unsurprisingly, a regular perturbation expansion in powers of $\nu$ breaks down, since it is not possible to impose the necessary boundary conditions for the current. This singularity of the elastic limit was to be expected on a physical basis, as it is not possible to obtain the behavior of weakly dissipative systems ($\nu\ll 1$) as a correction around the conservative case $\nu=0$, for which $\rho(x)=T$ and $j(x)=0$. Therefore,  a singular perturbation analysis should be done, looking for a suitable rescaling of the variables for $\nu\ll 1$. Equation (\ref{1.4e}) for the averages implies that
\begin{equation}\label{1.20}
    y_{\text{av}}=R(T)+{\cal O}(\nu), \quad j_{\text{av}}=-\nu R(T) x+{\cal O}(\nu^2), \quad \nu\ll 1.
\end{equation}
The average current vanishes linearly in $\nu$ in the limit $\nu\to 0^+$, as expected. Moreover, the average dissipation, obtained by combining eqs. (\ref{1.16b}) and (\ref{1.20}), is given by
\begin{equation}\label{1.26}
    d_{\text{av}}=\nu R(T)+{\cal O}(\nu^2).
\end{equation}
Therefore, it is sensible to propose the following rescaling of variables
\begin{equation}\label{1.21}
    j(x)=\nu \psi(x) , \quad p_j(x)=\frac{\Pi_\psi(x)}{\nu},
\end{equation}
which is consistent with the canonical equations (\ref{1.14}), since
\begin{subequations}\label{1.22}
\begin{eqnarray}
    \psi'&=&\frac{1}{\nu}j'=\frac{1}{\nu} \frac{\partial\cal H}{\partial p_j}=\frac{\partial\cal H}{\partial \Pi_\psi},\\
    \Pi'_\psi&=&\nu p_j'=-\nu\frac{\partial\cal H}{\partial j}=-\frac{\partial\cal H}{\partial\psi},
\end{eqnarray}
\end{subequations}
with the same Hamiltonian $\cal H$. In other words, eq. (\ref{1.21}) defines a ``canonical transformation'' from the pair of canonical conjugate variables $\{j,p_j\}$ to $\{\psi,\Pi_\psi\}$, a transformation that heals the singular behavior in the $\nu\to 0^+$ limit. The Hamiltonian can be now written as
\begin{equation}\label{1.23}
    {\cal H}=\frac{1}{2}Q(y) p_y^2-y \Pi_\psi-\nu {\hat{D}(y)}^{-1}\psi p_y
\end{equation}
in the rescaled variables. Notice that the transformation introduced is essential to obtain the correct ``dominant balance'' \cite{Be99} to the lowest order. In particular, before the rescaling, the term proportional to $y p_j$ was of the order of $\nu$ and the term proportional to $j p_y$ was of the order of unity; after the rescaling the orders of magnitude are interchanged, the term proportional to $y \Pi_\psi$ is of the order of unity while the term proportional to $\psi p_y$ is of the order of $\nu$. We now start from the zero-th order rescaled Hamiltonian by putting $\nu=0$ in eq. (\ref{1.23}),
\begin{equation}\label{1.24}
    {\cal H}_0=\frac{1}{2}Q(y)p_{y}^2-y \Pi_{\text{$\psi$}}.
\end{equation}
from which we we arrive at
\begin{subequations}\label{1.25}
\begin{eqnarray}
    y'&=&\frac{\partial{\cal H}_0}{\partial p_{y}}=Q(y) p_{y} \, ,  \label{1.25a} \\
    p'_{y}&=&-\frac{\partial{\cal H}_0}{\partial y}=-\frac{1}{2}\frac{dQ(y)}{dy}p_{y}^2+\Pi_{\text{$\psi$}} \, , \\
    \psi'&=&\frac{\partial{\cal H}_0}{\partial\Pi_{\text{$\psi$}}}=-y \, , \label{1.25b}\\
    \Pi'_{\text{$\psi$}}&=&-\frac{\partial{\cal H}_0}{\partial\psi}=0 \, . \label{1.25c}
\end{eqnarray}
\end{subequations}
In order not to clutter our formulas, we do not introduce a different notation for the canonical variables, although the approximate canonical equations (with ${\cal H}_0$) are different from the exact ones (with $\cal H$). We have only to remember that our results are valid to the lowest order in $\nu$. The canonical equations (\ref{1.25}) have to be solved with the boundary conditions
\begin{equation}\label{1.25c}
 y(\pm 1/2)=R(T), \qquad \psi(-1/2)=-\psi(1/2)=\Delta/2,
\end{equation}
where
\begin{equation}\label{1.25d}
\Delta=\frac{d}{\nu}=R(T) \frac{d}{d_{\text{av}}}
\end{equation}
is assumed to be of the order of unity, that is, $d=\mathcal{O}(\nu)$ or $d/d_{\text{av}}=\mathcal{O}(1)$. Thus, our rescaling allows us to obtain a solution for the optimal profiles for the density $\rho$, by inverting the relation $y=R(\rho)$, and the current $j=\nu\psi$, for integrated dissipations $d$ very different from its average value $d_{\text{av}}$.

It is worth noticing that $\psi$ is  a cyclic variable and its conjugate momentum is thus constant, $\Pi_\psi\equiv \Pi_{\psi 0}=\text{const.}$, see Eq. (\ref{1.25c}); this fact allows us to obtain a closed first order differential equation for $y(x)$ in the $\nu\ll 1$ limit. Moreover, by recalling Eq. (\ref{var12}), we have that
\begin{equation}\label{var13}
  \Pi_{\psi 0}=\frac{\partial G}{\partial\Delta},
\end{equation}
which gives the physical interpretation of this first integral of the approximate canonical equations: it is the partial derivative of the LDF with respect to the rescaled dissipation. The Hamiltonian ${\cal H}_0$ is also constant, since it does not depend explicitly on $x$, and combining (\ref{1.24}) and (\ref{1.25a}),
\begin{equation}\label{1.26b}
    {y'}^2=2Q(y) ({\cal H}_0+y \Pi_{\psi 0}), \qquad y(\pm 1/2)=R(T).
\end{equation}
Once this is solved, the rescaled current $\psi$ can be obtained from (\ref{1.25b})
\begin{equation}\label{1.27}
    \psi'=-y, \qquad \psi(-1/2)=-\psi(1/2)=R(T)\frac{d}{2d_{\text{av}}},
\end{equation}
so that the two constants $\mathcal{H}_0$ and $\Pi_{\psi 0}$ will be given in terms of the temperature $T$ and $d/d_{\text{av}}$. There are no more constants to be adjusted in the solution of eqs. (\ref{1.26b}) and (\ref{1.27}) due to the parity properties of $(y,\psi)$: $y$ is and even function of $x$ and $\psi$ is an odd function of $x$ in the interval $[-1/2,1/2]$. Of course, the average profiles $y_{\text{av}}(x)$ and $j_{\text{av}}(x)$ are reobtained from the canonical equations by putting $\Pi_\psi=0$ and $p_y=0$ therein. Equation (\ref{1.24}) implies that $\mathcal{H}_0=0$ over the average profiles.

The simple form of the differential equation (\ref{1.26b}) allows us to infer some of the properties of the optimal profile $y(x)$ associated to a given dissipation fluctuation in the limit of weakly-dissipative systems. First, notice that in general the solution of eq. (\ref{1.26b}) will be non-monotonic, exhibiting extrema in the interval $x\in[-\frac{1}{2},\frac{1}{2}]$. Moreover, taking into account that the function $Q(y)$ is positive defined, it follows that the profile at the extrema will take an unique value
\begin{equation}\label{1.27a}
  y_0 \equiv -\frac{\mathcal{H}_0}{\Pi_{\psi 0}}
\end{equation}
Note that $\Pi_{\psi 0}\neq 0$ for $d\neq d_{\text{av}}$ and, moreover, it must have a different sign that ${\cal H}_0$, i.e. $\sgn(\Pi_{\psi 0})\neq \sgn({H}_0)$, since $y(x)>0 \,\, \forall x$. Therefore, the optimal profile $y(x)$ can only have a single extremum (minimum or maximum) \cite{foot2}, that is located at $x=0$ because of symmetry reasons. By rewriting eq. (\ref{1.26b}) as
\begin{equation}\label{1.27b}
  y'^2=2 Q(y) \mathcal{H}_0 \left( 1-\frac{y}{y_0} \right)
\end{equation}
we conclude that the constant $\mathcal{H}_0$ and $y_0-y(x)$ must have the same sign $\forall x\in[-\frac{1}{2},\frac{1}{2}]$. Thus, for $\mathcal{H}_0>0$ the profile $y(x)$ has a single maximum, $y(x)>y(\pm 1/2)=R(T) \,\, \forall x$, and thus $d>d_{\text{av}}$. On the other hand, $\mathcal{H}_0<0$ implies a single minimum, $y(x)<R(T) \,\, \forall x$ and $d<d_{\text{av}}$. All these properties are confirmed below for particular examples, both analytically and numerically.

Interestingly, the leading behavior for the LDF can be also easily obtained in terms of the first integrals $\mathcal{H}_0$ and $\Pi_{\psi 0}$. In fact
\begin{equation}\label{1.28}
    G(d)\sim- \frac{1}{2}\int_{-1/2}^{1/2} dx\,Q(y) p_{y}^2=-\frac{1}{2}\int_{-1/2}^{1/2} dx\, \frac{{y'}^2}{Q(y)} ,
\end{equation}
and making use of eq. (\ref{1.26b}),
\begin{eqnarray}
    G(d) & \sim & - \left(\mathcal{H}_0+\Pi_{\psi 0} \int_{-1/2}^{1/2} dx \, y(x) \right) \nonumber \\
            &   =   & -\left(\mathcal{H}_0+\Pi_{\psi 0} \Delta\right) = - \mathcal{H}_0 \left(1-\frac{\Delta}{y_0}\right) \, . \label{1.29}
\end{eqnarray}
Thus, the remaining task consists in writing the constants $\mathcal{H}_0$ and $y_0$ (or equivalently $\mathcal{H}_0$ and $\Pi_{\psi 0}$) in terms of the integrated dissipation $d$ and the temperature at the boundaries $T$. Once this is done, the LDF follows from the simple expression given by  eq. (\ref{1.29}). Furthermore, we may obtain bounds for the profile extremum $y_0$ by taking into account that $G(d)<0$ for $d\neq d_{\text{av}}$: (i) for $\mathcal{H}_0>0$ we already know that $y_0$ is a maximum and this implies that  $y_0 > \Delta$, while (ii) for $\mathcal{H}_0<0$ we already know that $y_0$ corresponds to a minimum and thus $y_0 < \Delta$. Interestingly, we can also use Eq. (\ref{1.29}) together with Eq. (\ref{var13}) to obtain a simple relation between $\calH_0$, $G$ and $\partial G/\partial d$, namely
\begin{equation}\label{1.29b}
  -\calH_0=G+\Delta \frac{\partial G}{\partial\Delta}=G(d)+d \frac{\partial G(d)}{\partial d}=\text{const.} \,
\end{equation}
It must be stressed that this relation only holds in the weakly dissipative system limit, in the sense that the macroscopic dissipation coefficient is small, $\nu\ll 1$, and the considered integrated dissipation verifies that $d=\mathcal{O}(\nu)$.

For many systems of interest, the function $Q(y)$ is typically a homogeneous function of $y$ of degree $\gamma$,
\begin{equation}\label{1.30}
    Q(c y)=c^\gamma Q(y),
\end{equation}
where $c$ is an arbitrary real number, that is, $Q(y)\propto y^\gamma$. This type of dependence is common to many driven dissipative media, as for instance the general family of models that we will study in section \ref{s2} \cite{PLyH11a,PLyH12a} or different reaction-diffusion systems \cite{lutsko}. However, it should be noted that not all systems obey this homogeneity condition, e.g. symmetric simple exclusion processes with dissipative dynamics have a non-homogeneous $Q(y)$ \cite{Bodineau,Derrida}.  By introducing the scaling
\begin{equation}\label{1.31}
   y(x)= R(T) Y(x), \quad y_0=R(T) Y_0, \quad \mathcal{H}_0= R(T)^{2-\gamma} \widetilde{\mathcal{H}},
\end{equation}
Eq. (\ref{1.26b}) is transformed into
\begin{equation}\label{1.32}
    {Y'(x)}^2=2 \widetilde{\calH} Q(Y) \left(1-\frac{Y(x)}{Y_0}\right), \quad Y(\pm 1/2)=1.
\end{equation}
This is quite a natural transformation: we scale the variable $y$ with its value $R(T)$ at the boundaries; besides, we will be able to find a physically-relevant scaling variable for the dissipation LDF. Accordingly, we also introduce
\begin{subequations}\label{1.33}
\begin{eqnarray}
    &&\psi(x)=R(T) \Psi(x), \quad \Psi'(x)=-Y(x), \\ &&\Psi(-1/2)=\Psi(1/2)=\frac{\Delta}{2R(T)}=\frac{d}{2 d_{\text{av}}}.
\end{eqnarray}
\end{subequations}
The (even) solution of Eq. (\ref{1.32}) has the form
\begin{equation}\label{1.33b}
  Y=Y(x,\widetilde{\mathcal{H}},Y_0).
\end{equation}
As said before, there are no more integration constants when solving Eqs. (\ref{1.32})-(\ref{1.33}), since $Y$ (resp. $\Psi$) is an even (resp. odd) function of $x$. The boundary condition is
\begin{equation}\label{1.34}
  Y(x=1/2,\widetilde{\mathcal{H}},Y_0)=1 ,
\end{equation}
which implies that $Y_0=Y_0(\widetilde{\mathcal{H}})$, the scaled height $Y_0$ is only a function of the scaled Hamiltonian $\widetilde{\mathcal{H}}$. Now, taking into account that the (odd) solution of Eq. (\ref{1.33}) has the form
\begin{equation}\label{1.34b}
  \Psi=\Psi(x,\widetilde{\calH},Y_0),
\end{equation}
we have that
\begin{equation}\label{1.35}
  \Psi(x=-1/2,\widetilde{\mathcal{H}},Y_0(\widetilde{\mathcal{H}}))=\frac{d}{2 d_{\text{av}}}.
\end{equation}
Therefore,
\begin{equation}\label{1.36}
 \widetilde{\calH}=\widetilde{\calH}\left(\frac{d}{d_{\text{av}}}\right),
\end{equation}
that is, $\widetilde{\calH}$ is a function only of the integrated dissipation $d$ relative to its average value $d_{\text{av}}$.

This observation will be used in what follows to find a simple scaling form for the dissipation LDF. In fact, equation (\ref{1.29}) can be readily rewritten as
\begin{equation}\label{1.39}
    R(T)^{\gamma-2} G(d)\sim -  \left[ 1-\frac{d/d_{\text{av}}}{Y_0(\widetilde{\calH})}\right] \widetilde{\mathcal{H}}
\end{equation}
The equation above gives the general scaling of the LDF in the limit of weakly-dissipative systems: since both $\widetilde{\calH}$ and $Y_0$ are only  functions of $d/d_{\text{av}}$,
\begin{equation}\label{1.41}
    \left(\frac{d_{\text{av}}}{\nu}\right)^{\gamma-2}G(d)=- \left[ 1-\frac{d/d_{\text{av}}}{Y_0(\widetilde{\calH})}\right] \widetilde{\mathcal{H}}
\end{equation}
is only a function of $d/d_{\text{av}}$ [we have made use of Eq. (\ref{1.26}) for $R(T)$]. This is quite a strong result: it means that, for each value of $\gamma$, all the curves of $(d/d_{\text{av}})^{\gamma-2}G(d)$ plotted as a function of $d/d_{\text{av}}$ fall on a certain ``master'' curve for all values of the dissipation coefficient $\nu$, provided that $\nu\ll 1$ so we are dealing with a weakly dissipative system. The only hypothesis is that $Q(y)$ must be a homogeneous function of $y$, with an arbitrary degree $\gamma$, a rather general assumption satisfied in many cases of interest (see below). That being said, it is important to stress that the differential equation (\ref{1.32}) for $Y(x)$ contains the function $Q(y)$, that has $\gamma$ as a parameter. Thus, both $(Y,\Psi)$ and the rhs of eq. (\ref{1.41}) also contain $\gamma$ as a parameter; in principle, different physical models with different functions $Q(y)$ have different scaling functions. The simplest situation appears for $\gamma=2$, in that case Eq. (\ref{1.41}) predicts that $G(d)$ is only a function of the relative dissipation $d/d_{\text{av}}$, with no additional dependence on $\nu$.

Finally, it is also interesting to note that the optimal profiles $y$ and $\psi$ also have simple scaling forms. In fact, eq. (\ref{1.33b}), together with Eq. (\ref{1.31}), implies that
\begin{equation}\label{1.42}
    y(x)=\frac{d_{\text{av}}}{\nu} Y(x,\widetilde{\mathcal{H}},Y_0(\widetilde{\mathcal{H}})).
\end{equation}
On the other hand, eq. (\ref{1.34b}),  together with Eqs. (\ref{1.32}), yields that
\begin{equation}\label{1.43}
     \psi(x)=\frac{d_{\text{av}}}{\nu}
     \Psi(x,\widetilde{\mathcal{H}},Y_0(\widetilde{\mathcal{H}})).
\end{equation}
Therefore, both $y(x)$ and $\psi(x)$ multiplied by $\nu/d_{\text{av}}$ (that is, divided by $R(T)$) plotted as a function of $x$ collapse onto a single curve for constant $d/d_{\text{av}}$ for each value of $\gamma$ and all possible $\nu\ll 1$ in the weakly dissipative regime. From eqs. (\ref{1.42}) and (\ref{1.43}), the optimal profiles for the density and the current are readily obtained, since $y=R(\rho)$ and $j=\nu\psi$. On the other hand, the first integrals $\calH_0$ and $\Pi_{\psi 0}$ follow from Eqs. (\ref{1.27a}) and (\ref{1.31}),
\begin{equation}\label{1.44}
  \calH_0=\left(\frac{d_{\text{av}}}{\nu}\right)^{2-\gamma} \, \widetilde{\calH}, \quad \Pi_{\psi 0}=-\left(\frac{d_{\text{av}}}{\nu}\right)^{1-\gamma} \frac{\widetilde{\calH}}{Y_0(\widetilde{\calH})}.
\end{equation}

\subsection{Strongly-dissipative systems, $\nu\gg 1$}
\label{s1.5c}

We now proceed to analyze the limit $\nu\gg 1$, i.e. the limit of strongly-dissipative dynamics. Equation (\ref{1.4h}) for the average profile implies that $y_{\text{av}}(x)$ develops two boundary layers of width $\ell_{\nu}\sim \nu^{-1/2}$ close to $x=\pm 1/2$, where most of the system energy is localized. This is reasonable on physical grounds: for $\nu\gg 1$ one expects that the injection of energy through the boundaries would be limited to a small region near them: most of the energy has been dissipated before reaching the bulk of the system, effectively decoupling the system into two almost-independent halves. This picture can be used to simplify the integration of the system of canonical equations (\ref{1.14}):  we can restrict ourselves to the half interval $x\in[-1/2,0]$ and use the boundary conditions
\begin{equation}\label{1.50}
   y(-1/2)=R(T),  \; j(-1/2)=d/2, \; y'(0)=0, \;j(0)=0,
\end{equation}
because of the symmetry of the solutions ($y$ even, $j$ odd). Now, by introducing the following rescaling (suggested by the typical lengthscale $\ell_{\nu}\sim \nu^{-1/2}$),
\begin{equation}\label{1.51}
  j=\sqrt{\nu} \psi, \quad X=\sqrt{\nu} \left(x+\frac{1}{2}\right), \quad p_y=\sqrt{\nu} \Pi_y,
\end{equation}
we arrive at the equivalent system of equations
\begin{subequations}\label{1.52}
\begin{equation}\label{1.52a}
     \frac{dy}{dX}=Q(y)\Pi_y-{\hat{D}}^{-1}(y)\psi,
\end{equation}
\begin{equation}\label{1.52b}
    \frac{d\psi}{dX}= -y,
\end{equation}
\begin{equation}\label{1.52c}
    \frac{d\Pi_y}{dX}=-\frac{dQ(y)}{dy}  \frac{\Pi_y^2}{2}+\frac{d{\hat{D}}^{-1}(y)}{dy}\psi \Pi_y+p_j,
\end{equation}
\begin{equation}\label{1.52d}
    \frac{dp_j}{dX}={\hat{D}}^{-1}(y)\Pi_y
\end{equation}
\end{subequations}
with the boundary conditions
\begin{subequations}\label{1.53}
    \begin{eqnarray}
  y(X=0)=R(T), &\quad& \left.\frac{dy}{dX}\right|_{X=\sqrt{\nu}/2}=0, \\ \psi(X=0)=\frac{d}{2\sqrt{\nu}}, &\quad& \psi(X=\sqrt{\nu}/2)=0.
    \end{eqnarray}
\end{subequations}
Interestingly, $\nu$ does not appear explicitly in the rescaled canonical equations (\ref{1.52}), but only in the boundary conditions. Therefore, in the limit $\nu\to\infty$ we have to solve (\ref{1.52}) with the boundary conditions
\begin{subequations}\label{1.54}
    \begin{eqnarray}
     y(X=0)=R(T), &\quad&  \psi(X=0)=\tilde{d}, \\
     \lim_{X\to\infty} \frac{dy}{dX}=0, &\quad& \lim_{X\to\infty} \psi=0.
    \end{eqnarray}
\end{subequations}
where we have defined
\begin{equation}\label{1.54b}
  \tilde{d}=\frac{d}{2\sqrt{\nu}},
\end{equation}
which is assumed to be of the order of unity. In fact, from Eq. (\ref{1.8c}) one gets
\begin{equation}\label{1.54c}
  d=2\nu \int_{-1/2}^0 dx y(x)=2 \sqrt{\nu} \int_0^{\sqrt{\nu}/2} dX\, y(X),
\end{equation}
that is,
\begin{equation}\label{1.54d}
  \tilde{d}\sim \int_0^{\infty} dX \, y(X), \quad \nu\gg 1.
\end{equation}
In this strongly-dissipative regime, the canonical equations themselves are not simplified, but a physically appealing picture emerges: the system decouples in two independent boundary shells of width $\mathcal{O}(\sqrt{\nu})$ close to the boundaries, where the rescaled variable $X=\mathcal{O}(1)$ (we have restricted ourselves to the semi-interval [-1/2,0], the solution in [0,1/2] is found by the symmetry arguments already used). Moreover,  a simple scaling can be derived for the LDF $G(d)$, Eq. (\ref{1.15d}),
\begin{equation}\label{1.55}
  G(d)=-\int_{-1/2}^0 dx Q(y) p_y^2\sim -\sqrt{\nu} \int_0^\infty dX Q(y) \Pi_y^2,
\end{equation}
where $y$ and $\Pi_y$ are the solutions of (\ref{1.52}) with the boundary conditions (\ref{1.54}). Therefore, both $y$ and $\Pi_y$ depend on $R(T)$ and $\tilde{d}$ through  the boundary conditions and $G(d)=\sqrt{\nu} F(R(T),\tilde{d})$ where $F$ is a certain function.

A particularly simple situation appear when both the mobility $\sigma(y)$ and the diffusivity  $D(y)$ are proportional to some power of $y$, so that (i) the function $Q(y)$ is homogeneous, $Q(y)\propto y^\gamma$, as in Eq. (\ref{1.30}), and (ii) the effective diffusivity $\hat{D}$ does not depend on $y$, $\hat{D}(y)=\hat{D}=\text{const.}$ as discussed in Sec. \ref{s1}.  In fact, this is the case for the general class of dissipative models analyzed in Sec. \ref{s2}. The average dissipation for $\nu\gg 1$ is
\begin{equation}\label{1.55b}
  d_{\text{av}}\sim2\sqrt{\nu \hat{D}} R(T) \; \Rightarrow \; \tilde{d}=\sqrt{\hat{D}}R(T) \frac{d}{d_{\text{av}}}.
\end{equation}
The canonical equations (\ref{1.52}) can be analyzed following a line of reasoning similar to the one used in the weakly dissipative system limit. Since the details are not necessary for the work presented here, we only give the final result for the LDF, that is
\begin{equation}\label{1.56}
  G(d)=-\sqrt{\frac{\nu}{\hat{D}}} \, \mathcal{F}\left(\left[R(T)\right]^{\gamma-2},\frac{d}{d_{\text{av}}}\right),
\end{equation}
where $\mathcal{F}$ is a certain scaling function. The scaling in Eq. (\ref{1.56}) is more complex than in the weakly dissipative system limit. For instance, in the case $\gamma=2$ we get that $G(d)=\sqrt{\nu/\hat{D}}\, \mathcal{F}(1,d/d_{\text{av}})$, so the LDF curves, once rescaled by $(\hat{D}/\nu)^{1/2}$, collapse for all $\nu\gg 1$ when plotted as a function of the relative dissipation $d/d_{\text{av}}$. The factor $\sqrt{\nu}$ in front of the scaling function accounts for the strong supression of the fluctuations of the dissipation that takes place in strongly-dissipative systems: for a given value of the relative dissipation $d/d_{\text{av}}$, the probability of such a fluctuation decreases exponentially with $\sqrt{\nu}$.

\section{A general class of nonlinear driven dissipative models}
\label{s2}

In order to investigate in detail the validity of the general framework presented in previous sections, we now introduce a broad class of dissipative lattice models with stochastic microscopic dynamics that contain the essential ingredients characterizing many dissipative media, namely: (i) nonlinear diffusive dynamics, (ii) bulk dissipation, and (iii) boundary driving. For the sake of simplicity, we will present them for the one-dimensional (1D) case, but the extension to arbitrary dimension is straightforward.

We thus consider a system defined on a 1D lattice with $N$ sites. A configuration at a given time step $p$ is given by $\bm{\rho}=\{\rho_{l,p}\}$, $l=1,\ldots,N$, where $\rho_{l,p} \geq 0$ is the \emph{energy} of the $l$-th site at time $p$, so the total energy of the system at this time is $E_p=\sum_{l=1}^N \rho_{l,p}$, see Fig. \ref{sketch}. The dynamics is stochastic and sequential, and proceeds via collisions between nearest neighbors. In an elementary step, a nearest neighbor pair of sites $(l,l+1)$ interacts with probability
\begin{equation}\label{2.0}
   P_{l,p}(\bm{\rho})=\frac{f(\Sigma_{l,p})}{\sum_{l'=1}^L f(\Sigma_{l',p})}, \quad \Sigma_{l,p}=\rho_{l,p}+\rho_{l+1,p},
\end{equation}
where $f$ is a given function of the pair energy $\Sigma_{l,p}$, and $L$ is the number of possible pairs. Clearly $L\sim N$, but the particular relation depends on the boundary conditions imposed (e.g., $L=N+1$ for open boundaries while $L=N$ for the periodic case). Once a pair is chosen, a certain fraction of its energy, namely  $(1-\alpha)\Sigma_{l,p}$, is dissipated to the environment, mimicking the energy drain observed in real dissipative media. The remaining energy $\alpha \Sigma_{l,p}$ is then randomly redistributed between both sites,
\begin{equation}\label{2.1}
    \rho_{l,p+1}=z_p \alpha \Sigma_{l,p} \, , \quad \rho_{l+1,p+1}=(1-z_p) \alpha \Sigma_{l,p} \,,
\end{equation}
with $z_p$ an homogeneously distributed random number in the interval $[0,1]$. This microscopic random exchange mechanism yields nonlinear diffusion at the mesoscale, being an accurate representation of the coarse-grained local energy dynamics in many dissipative systems. The above dynamics defines the evolution of all bulk pairs, $l=1,\ldots,N-1$. In addition, and depending on the boundary conditions imposed, boundary sites might interact with thermal baths at both ends, possibly at different temperatures $T_L$ (left) and $T_R$ (right). In this case the dynamics is
\begin{equation}\label{2.2}
    \rho_{1,p+1}=z_p \alpha (e_{1,p}+\widetilde{e}_{L}), \qquad \rho_{N,p+1}=z_p \alpha (e_{N,p}+\widetilde{e}_{R}),
\end{equation}
when the first (last) site interacts with its neighboring thermal reservoir. Here $\widetilde{e}_{k}$, $k=L,R$, is an energy randomly drawn at each step from the canonical distribution at temperature $T_{k}$, that is, with probability $\text{prob}(\widetilde{e}_{k})=T_{k}^{-1} \exp (-\widetilde{e}_{k}/T_{k})$ (our unit of temperature is fixed by making $k_B=1$), see Fig. \ref{sketch}. We may consider instead an isolated system with periodic boundary conditions, such that $L=N$ and Eqs. (\ref{2.0}) and (\ref{2.1}) remain valid for $l=0$ ($l=N$) with the substitution $\rho_{0,p}=\rho_{N,p}$ ($\rho_{N+1,p}=\rho_{1,p}$).

\begin{figure}
\vspace{0.5cm}
\centerline{
\includegraphics[width=8cm]{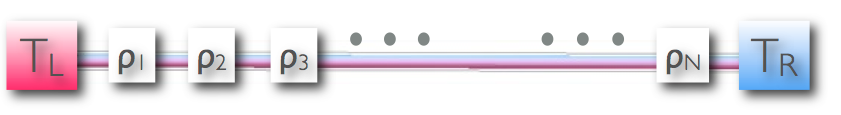}
}
\caption{\small The model is defined on lattice sites, each one characterized by an energy $\rho_l$. The dynamics is stochastic and proceeds via random collisions between nearest neighbors where part of the pair energy is dissipated to the environment and the rest is randomly redistributed within the pair. Such dynamics mimics at the mesoscopic level the evolution of a wide class of systems characterized by a nontrivial competition between diffusion and dissipation.}
\label{sketch_model}
\end{figure}

The simplest dynamics corresponds to  $f(\Sigma_{l,p})=1$ in Eq. (\ref{2.0}). In this case all (nearest neighbor) pairs collide with equal probability $P_{l,p}=L^{-1}$, independently of their energy. This choice (together with $\alpha=1$ above) corresponds to the Kipnis-Marchioro-Presutti (KMP) model of energy transport \cite{kmp}, which can be considered as a coarse-grained description of the physics of a large class of quasi-1D real diffusive systems. For instance, it is one of the very few instances where Fourier's law can be rigorously proved \cite{kmp}. In addition, the KMP model has been used to investigate the validity of the additivity principle for current fluctuations \cite{BD} and the Gallavotti-Cohen fluctuation theorem \cite{GC} and its generalization in refs. \cite{Pablo,iso}. Another simple, but physically relevant, choice is $f(\Sigma_{l,p})=\Sigma_{l,p}$, so that $P_{l,p}\sim\Sigma_{l,p}/(2E_p)$ for a large system. A variant of this model has been recently used to study compact wave propagation in microscopic nonlinear diffusion \cite{HyK11}. In general, the models here introduced can be regarded as a \textit{toy} description of dense granular gases: particles cannot freely move but may collide with their nearest neighbors, losing a fraction of the pair energy and exchanging the rest thereof randomly. The inelasticity parameter can be thus considered as the analogue to the restitution coefficient in granular systems \cite{PyL01}: energy is conserved in the dynamics only for $\alpha=1$, while it is continuously dissipated for any $0\leq \alpha<1$. Thus, in an isolated system (without boundary driving) the energy would decrease monotonically in time. However, if energy is injected, for instance via coupling to boundary thermal baths as described above, a steady state will be eventually reached where energy injection and dissipation balance each other. The class of models here presented is an optimal candidate to study dissipation statistics because: (a) one can obtain explicit predictions for the LDF, and (b) its simple dynamical rules allow for a detailed numerical study. The chances are that our results remain valid for more complex dissipative media with similar macroscopic dynamics.

The hydrodynamic evolution laws for this family of models have been recently studied in detail in ref. \cite{PLyH12a}. In the large system size limit, both continuous space and time variables can be introduced, as well as the relevant hydrodynamic fields: energy density $\rho(x,t)$, current $j(x,t)$ and dissipation $d(x,t)$. This mesoscopic description is expected to be valid not only for this particular class of models but for many driven dissipative media over a certain ÒhydrodynamicÓ time scale, much larger than the one characteristic of the microscopic dynamics. Over the fast (microscopic) time scale, the system forgets the initial conditions and relaxes to a local equilibrium state in which all the properties of the system become functionals of the hydrodynamic fields. Afterwards, over the much slower macroscopic time scale, the system eventually approaches the steady state following certain hydrodynamic law. For the family of models introduced in this paper, the time evolution of the energy density can be shown to obey a fluctuating balance equation of the general form \cite{PLyH12a}
\be
\partial_t\rho(x,t)=-\partial_x j(x,t)  + d(x,t) \, , \nonumber
\ee
which is just the starting point for the generalization of Macroscopic Fluctuation Theory to dissipative systems developed in previous sections. The first term in the rhs of this equation accounts for the diffusive spreading of the energy, and it is also present in the conservative case, while the second one gives the rate of energy dissipation in the bulk. It is important to note here that the models microscopic dynamics must be quasi-elastic (with $(1-\alpha)\sim L^{-2}$, see eq. (\ref{alfa}) below) in order to ensure that both diffusion and dissipation take place over the same time scale in the continuum limit. Using a local equilibrium approximation, the current and dissipation fields can be expressed as functions of the local energy density \cite{PLyH12a}. In particular, the fluctuating current can be written as
\be
j(x,t)=-D(\rho)\partial_x \rho + \xi \, ,
\label{fou}
\ee
where the first term is nothing but Fourier's (equivalently Fick's) law with a diffusivity $D(\rho)$. This transport coefficient can be explicitely calculated for the general family of models here presented \cite{PLyH12a}, obtaining
\be
D(\rho)=\frac{1}{6}\int_0^\infty dr\, r^7 f(\rho r^2) e^{-r^2} \, ,
\label{diffu}
\ee
where $f(\Sigma)$ is the function defining the microscopic collision rate, see eq. (\ref{2.0}). On the other hand, the second term $\xi$ in eq. (\ref{fou}) is a noise perturbation, white and gaussian. These gaussian fluctuations are expected to emerge for most situations in the appropriate mesoscopic limit as a result of a central limit theorem. Microscopic interactions can be highly complicated, but the ensuing fluctuations of the slow hydrodynamic fields result from the sum of an enormous amount of random events at the microscale which give rise to Gaussian statistics at the mesoscale. In the present case, a proof of the gaussian character of the noise can be given, due to the simplicity of the class of models considered \cite{PLyH12a}. The current noise amplitude is $\sigma(\rho)/L$ (that is, the noise strength scales as $L^{-1/2}$), where $\sigma(\rho)$ is often referred to as the mobility in the literature. This coefficient can be again explicitely computed within the local equilibrium approximation,
\be
\sigma(\rho)=\frac{\rho^2}{3} \int_0^\infty dr \, r^7 f(\rho  r^2) e^{-r^2} \, .
\label{sigma}
\ee
Remarkably, a direct inspection of eqs. (\ref{diffu}) and (\ref{sigma}) reveals a simple relation between mobility and diffusivity
\be
\sigma(\rho)=2\rho^2 D(\rho) \, ,
\label{fdr}
\ee
which is nothing but a general fluctuation-dissipation relation for the dissipative case. In fact, it is the same one as in the conservative case, because of the quasi-elasticity of the underlying microscopic (stochastic) dynamics, see eq. (\ref{alfa}). On the other hand, the dissipation field can be written as $d(x,t)=-\nu R(\rho)$, where $\nu$ is a macroscopic dissipation coefficient which can be related to the inelasticity of the underlying microscopic dynamics
\be
1-\alpha\equiv\frac{\nu}{2L^2} \, ,
\label{alfa}
\ee
and $R(\rho)$ is a new transport coefficient, absent in conservative systems
\be
R(\rho)=\rho \int_0^\infty dr r^5 f(\rho r^2) e^{-r^2}\, .
\label{dissip}
\ee
Interestingly, $R(\rho)$ can be related to the diffusivity. By differentiating eq. (\ref{dissip}) with respect to $\rho$ after a change of variables $z=r \sqrt{\rho}$, it is found that
\begin{equation}\label{2.26b}
  D(\rho)=\frac{1}{6}\frac{dR(\rho)}{d\rho}+\frac{R(\rho)}{3\rho}.
\end{equation}
In fact, given the diffusivity, this equation can be considered as a first order differential equation for $R(\rho)$. The solution thereof, with an appropriate boundary condition (normally $R(\rho=0)=0$), is equivalent to calculate the integral in eq. (\ref{dissip}).

It should be stressed that the dissipation field $d(x,t)$ has no intrinsic noise, so its observed local fluctuations are enslaved to those of the density. This stems from the subdominant role of the noise affecting the dissipative term: its strength scales as $L^{-3/2}$ as a consequence of the quasi-elasticity of the microscopic dynamics, see eq. (\ref{alfa}), so it is negligible against the current noise in the mesoscopic limit \cite{PLyH12a}.

\section{Large deviations of the dissipated energy}
\label{s4}

In what follows we will apply the theory developed in sections \ref{s1}-\ref{s1.5} to the general class of dissipative models introduced in section \ref{s2}.
More concretely, we will restrict ourselves to the following choice for the collision rate function
\begin{equation}\label{4.0}
  f(\rho)=\frac{2}{\Gamma(\beta+3)}\rho^\beta,
\end{equation}
that is, $f(\rho)\propto\rho^\beta$, with $\beta>-3$ but otherwise arbitrary.  We have introduced the constant $2/\Gamma(\beta+3)$ \cite{gamma} for the sake of convenience, as it simplifies the expressions of the transport coefficients, see below. For $\beta=0$, $f(\rho)=1$ and all the pairs collide with equal probability, independently of their energy value. Thus, the dissipative generalization of the KMP model introduced in ref. \cite{PLyH11a} is recovered. For $\beta=1$, $f(\rho)=\rho/3$ and the colliding pairs are chosen with probability proportional to their energy. The conservative case has been recently analyzed in \cite{HyK11}.

The transport coefficients for this family of models are easily calculated. The dissipation coefficient is readily obtained from Eq. (\ref{dissip}),
\begin{equation}\label{4.1}
    R(\rho)=\frac{2}{\Gamma(\beta+3)}\rho^{\beta+1}\int_0^\infty dr \, r^{5+2\beta} e^{-r^2}=\rho^{\beta+1},
\end{equation}
that gives the rationale behind the choice of the proportionality constant in Eq. (\ref{4.0}). The diffusivity is calculated by substituting Eq. (\ref{4.1}) into (\ref{2.26b}),
\begin{equation}\label{4.2}
    D(\rho)=\frac{\beta+3}{6}\rho^\beta.
\end{equation}
Of course, the same result is obtained with eq. (\ref{diffu}). Finally, the mobility $\sigma(\rho)$ follows from the fluctuation-dissipation relation (\ref{fdr}), which gives it in terms of the diffusivity,
\begin{equation}\label{4.3}
    \sigma(\rho)=2\rho^2 D(\rho)=\frac{\beta+3}{3} \rho^{\beta+2}.
\end{equation}
Of course, for $\beta=0$ the values of the transport coefficients of the dissipative version of the KMP model are recovered, $D(\rho)=1/2$, $\sigma(\rho)=\rho^2$, and $R(\rho)=\rho$ \cite{PLyH11a}. Interestingly, the algebraic dependence of transport coefficients with the energy density $\rho$ appears ubiquitously in real systems. One example is granular materials \cite{PyL01}, where the density field $\rho$ may be assimilated to the local granular temperature. For the hard sphere model, the average collision rate is proportional to the square root of the granular temperature. Thus, this granular gas case should correspond to $\beta=1/2$, and in fact it is found that $D(\rho)\propto\rho^{1/2}$ while the dissipative term goes as $R(\rho)\propto\rho^{3/2}$. The latter is responsible for the algebraic decay with time of the granular temperature (Haff's law; $\propto t^{-2}$ for large times) observed in the homogeneous case when the system is isolated \cite{PyL01}.

Before going into the details, it is convenient to write the explicit form of the auxiliary variable $y$, related to the density $\rho$, defined in eq.(\ref{1.4a}). For our family of models,
\begin{equation}\label{4.8}
    y=\rho^{1+\beta}, \quad \rho=y^{\frac{1}{1+\beta}},
\end{equation}
where we have made use of eq. (\ref{dissip}). Following the notation introduced in section \ref{s1}, we also need the ``effective'' diffusivity $\hat{D}(y)$, given by eq. (\ref{1.4f}), and the mobility $\hat{\sigma}=\sigma(y)$, both written in terms of $y$,
\begin{subequations}\label{4.11}
\begin{equation}\label{4.11a}
    \hat{D}(y)=D(\rho(y))\frac{d\rho}{dy}=\frac{3+\beta}{6(1+\beta)},
\end{equation}
\begin{equation}\label{4.11b}
    \hat{\sigma}(y)=\sigma(\rho(y))=\frac{3+\beta}{3}y^{\frac{2+\beta}{1+\beta}}.
\end{equation}
\end{subequations}
Hence, as anticipated in section \ref{s1}, while the ``true'' mobility $D(\rho)$ depends on $\rho$, see eq. (\ref{diffu}), the ``effective'' diffusivity  is constant, $\hat{D}(y)=\hat{D}$. This allows us to calculate explicitly the average profiles for the density and the current, using the linearity on $y$ of eq. (\ref{1.4h}) above. The steady average solution is thus
\begin{eqnarray}
    y_{\text{av}}(x)&=&T^{1+\beta}\, \frac{\cosh\left(x\sqrt{\frac{\nu}{\hat{D}}}\right)}{\cosh\sqrt{\frac{\nu}{4\hat{D}}}} \, , \label{4.12}\\
    j_{\text{av}}(x)&=&-\hat{D}y'_{\text{av}}=-T^{1+\beta}\sqrt{\nu \hat{D}}\, \frac{\sinh\left(x\sqrt{\frac{\nu}{\hat{D}}}\right)}{\cosh\sqrt{\frac{\nu}{4\hat{D}}}} \, .\label{4.12a}
\end{eqnarray}
Moreover, the average density is readily obtained by combining (\ref{4.8}) and (\ref{4.12}),
\begin{equation}\label{4.13}
    \rho_{\text{av}}(x)=T \left[\frac{\cosh\left(x\sqrt{\frac{\nu}{\hat{D}}}\right)}{\cosh\sqrt{\frac{\nu}{4\hat{D}}}}\right]^{\frac{1}{1+\beta}}\, .
\end{equation}
The validity of these hydrodynamic predictions has been tested via extensive numerical experiments in ref. \cite{PLyH12a}. Notice that one can define an unique natural lengthscale associated to a given $\nu$ from the hydrodynamic profiles above, namely $\ell_{\nu}=\sqrt{\hat{D}/\nu}$. This is the lengthscale over which profiles vary appreciably, and it decreases like $\sim \nu^{-1/2}$ as $\nu$ grows. In fact this observation suggests that, in the limit of strongly-dissipative systems $\nu\gg 1$, boundary energy layers develop localized around the thermal baths, effectively decoupling the system in two almost-independent parts, an observation that we have already used in Section \ref{s1.5c} to obtain a simple scaling relation for the dissipation LDF for strongly dissipative systems $\nu\gg 1$.

We are interested in the probability of a given fluctuation of the dissipated energy $d$, integrated over space and time, as defined in eq. (\ref{1.7}). As described in section \ref{s1}, this probability obeys a large deviation principle
\begin{equation}\label{4.13b}
    {\cal P}_\tau(d) \sim \exp \left[+\tau L \, G(d)\right],
\end{equation}
where the large deviation function $G(d)$ is obtained by solving a variational problem for the so-called ``optimal'' profiles $\{\rho_0(x;d),j_0(x;d)\}$ which sustain the considered fluctuation, after an (additivity) conjecture on the time-independence of optimal paths. In fact, making use of eq. (\ref{1.15d}),
\begin{equation}\label{4.14}
    G(d)=-\frac{1}{2}\int_{-1/2}^{1/2} dx \, Q(y) p_y^2,
\end{equation}
with $\{y(x),p_y(x)\}$ being the solutions of the canonical equations (\ref{1.14}) with the appropriate boundary conditions. For the family of models introduced in eq. (\ref{4.0}) for the collision probability, the Hamiltonian defined in eq. (\ref{1.13a}) is
\begin{equation}\label{4.14b}
    \mathcal{H}=\frac{1}{2}Q(y)p_y^2-\frac{1}{\hat{D}}j p_y-\nu y p_j.
\end{equation}
The auxiliary function $Q(y)$ in the equations above has been defined in eq. (\ref{1.13b}), for the case we are analyzing
\begin{equation}\label{4.16}
    Q(y)=\frac{\hat{\sigma}(y)}{{\hat{D}}^2}=\frac{12 (1+\beta)^2}{3+\beta} y^{\frac{2+\beta}{1+\beta}},
\end{equation}
that is, $Q(y)$ is basically the mobility, since $\hat{D}$ is constant, independent of $y$. Thus, $Q(y)$ is an homogeneous function of degree $\gamma$,
\begin{equation}\label{4.17}
    Q(y)=c \, y^\gamma, \quad \gamma=\frac{2+\beta}{1+\beta},
\end{equation}
with $c=12 (1+\beta)^2/(3+\beta)$. The parameter $\gamma$ varies from the value $\gamma=2$ for the case $\beta=0$, which corresponds to the dissipative KMP model, to $\gamma=1$ for the limit $\beta\to\infty$. The canonical equations determine the optimal profiles,
\begin{subequations}\label{4.15}
\begin{equation}\label{4.15a}
     y'=\frac{\partial\cal H}{\partial p_y}=Q(y)p_y-\frac{j}{\hat{D}},
\end{equation}
\begin{equation}\label{4.15b}
    j'=\frac{\partial\cal H}{\partial p_j}=-\nu y,
\end{equation}
\begin{equation}\label{4.15c}
    p'_y=-\frac{\partial\cal H}{\partial y}=-\frac{dQ(y)}{dy}  \frac{p_y^2}{2}+\nu p_j,
\end{equation}
\begin{equation}\label{4.15d}
    p'_j=-\frac{\partial\cal H}{\partial j}=\frac{p_y}{\hat{D}}.
\end{equation}
\end{subequations}
The boundary conditions for this system of differential equations are
\begin{equation}\label{4.15e}
    y(\pm 1/2)=T^{1+\beta}, \quad j(-1/2)=-j(1/2)=\frac{d}{2}.
\end{equation}
The main difference with the general system of canonical equations (\ref{1.14}) comes from $\hat{D}$ being constant and not a function of $y$.  Of course, just as in the general case, the canonical equations (\ref{4.15}) have a particular solution with vanishing canonical momenta, $p_y=0$, $p_j=0$,
\begin{equation}\label{4.18}
    y'=-\frac{j}{\hat{D}}, \quad j'=-\nu y,
\end{equation}
that is the particularization of eq. (\ref{1.4e}) for our family of models. The solution thereof are the average profiles (\ref{4.12})-(\ref{4.12a}).

\subsection{Typical fluctuations and gaussian behavior}
\label{s4a}

As described in subsection \ref{s1.5a}, gaussian fluctuations are expected for small deviations from the average dissipation behavior, $P_\tau(d) \approx \exp [ -L\tau (d-d_{\text{av}})^2/(2 d_{\text{av}}^2 \Lambda_{\nu}^2)]$, see eq. (\ref{1.19}). Following the general theory, small fluctuations correspond to small canonical momenta, since the average behavior is obtained for $p_y=0$, $p_j=0$. We linearize the canonical equations by introducing the parameter $\epsilon=(d-d_{\text{av}})/d_{\text{av}}\ll 1$. Recalling eq. (\ref{1.16b}), and making use of eq. (\ref{4.12}), the average dissipation is
\begin{equation}\label{4.18a}
    d_{\text{av}}=\nu \int_{-1/2}^{1/2} dx\, y_{\text{av}}(x)=T^{1+\beta} \sqrt{4\nu \hat{D}} \tanh \sqrt{\frac{\nu}{4\hat{D}}}.
\end{equation}
\begin{figure}
\centerline{
\includegraphics[width=9cm]{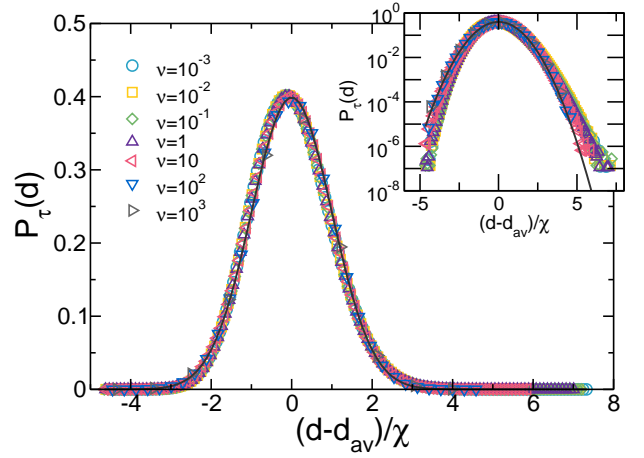}}
\vspace{-0.25cm}
\caption{(Color online) Probability distribution for the dissipated energy, integrated over space and a long time $\tau$, plotted versus the reduced variable $(d-d_{\text{av}})/\chi$ (of unit variance) for many different values of $\nu\in[10^{-3},10^3]$, for the case $\beta=0$. Inset: Semilog plot of the same data. In both cases the line is the normal distribution. Gaussian statistics is observed for typical fluctuations, but the tails already show signs of asymmetry.
}
\label{comparisontogaussian}
\end{figure}
Writing the canonical variables as their averages plus a linear correction in $\epsilon$, see eq. (\ref{1.16c}), we find at first order in $\epsilon$ the following set of equations
\begin{subequations}\label{4.19}
\begin{equation}\label{4.19a}
    \Delta y'=Q(y_{\text{av}})\Delta p_y-\frac{1}{\hat{D}} \Delta j,
\end{equation}
\begin{equation}\label{4.19b}
    \Delta j'=-\nu \Delta y,
\end{equation}
\begin{equation}\label{4.19c}
    \Delta p'_y=\nu \Delta p_j,
\end{equation}
\begin{equation}\label{4.19d}
    \Delta p'_j=\frac{1}{\hat{D}} \Delta p_y,
\end{equation}
\end{subequations}
which particularizes eq. (\ref{1.17}) for our family of models. The boundary conditions are $\Delta y(\pm 1/2)=0$, $\Delta j(-1/2)=-\Delta j(1/2)=d_{\text{av}}/2$. The solution of this system must be inserted into eq. (\ref{1.19b}), which gives the variance of the gaussian distribution, $\chi^2\equiv d_{\text{av}}^2 \Lambda_{\nu}^2/L\tau$.  The canonical momentum $\Delta p_y$ is directly obtained by integrating eqs. (\ref{4.19c}) and (\ref{4.19d}),
\begin{equation}\label{4.20}
    \Delta p_y=K \sinh\left(x\sqrt{\frac{\nu}{\hat{D}}}\right),
\end{equation}
since $\Delta p_y$ must be an odd function of $x$ as a consequence of $y$ being even. The constant $K$ is to be determined with the aid of the boundary conditions, but this can only be done after solving eqs. (\ref{4.19a}) and (\ref{4.19b}), having previously inserted (\ref{4.20}) into them. Substitution of eq. (\ref{4.20}) into eq. (\ref{1.19b}) gives
\begin{equation}\label{4.21}
    \Lambda_{\nu}^2= K^2 \left(\int_{-1/2}^{1/2} dx \, Q(y_{\text{av}}) \sinh^2\left(x\sqrt{\frac{\nu}{\hat{D}}}\right)\right)^{-1}.
\end{equation}
In order to evaluate the integral, eq. (\ref{4.17}) for  $Q(y)$ must be used, $Q(y)\propto y^\gamma$, with the parameter $\gamma$ being a function of $\beta$, $1<\gamma\leq 2$. We now analyze the simplest choice $\beta=0$, that is, $\gamma=2$, that corresponds to the dissipative version of the KMP model introduced in \cite{PLyH11a}. In this case, the calculation is straightforward and yields
\begin{equation}\label{4.22}
    \Lambda_{\nu}^2=\frac{\sinh(2\sqrt{2\nu}) - 2\sqrt{2\nu}}{4 \sqrt{2\nu} \sinh^2(\sqrt{2\nu})}.
\end{equation}
Interestingly, $\Lambda_{\nu}^2 \sim 1/3$ independent of $\nu$ in the limit of weakly-dissipative systems $\nu\ll 1$. This can be understood as a reminiscence of the scaling of $G(d)$ derived in section \ref{s1.5}. In fact, eq. (\ref{1.41}) tells us that, for $\gamma=2$, $G(d)$ is just a function of $d/d_{\text{av}}$. In the gaussian approximation, this implies the convergence of $\Lambda_{\nu}^2$ to a constant value in the quasi-elastic limit as $\nu\to 0^+$.  On the other hand, $\Lambda_{\nu}^2 \sim (2\sqrt{2\nu})^{-1}$ for $\nu\gg 1$, which is consistent with the suppression of dissipation fluctuations previously found in the strongly inelastic regime, as expressed by the general scaling of the LDF given by Eq. (\ref{1.56}). The same qualitative observations apply to other values of $\beta$, though the calculation is more convoluted.

\begin{figure}
\centerline{
\includegraphics[width=9cm]{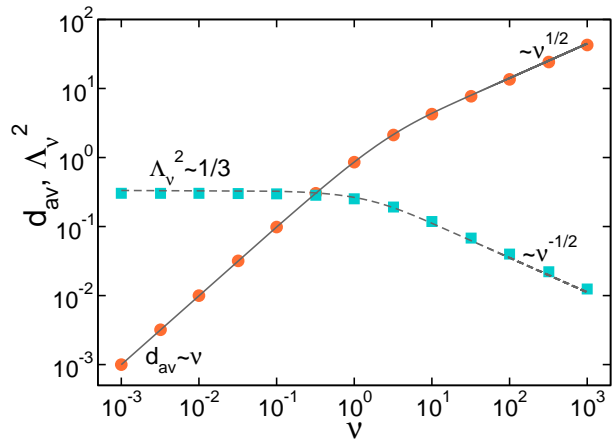}}
\vspace{-0.25cm}
\caption{(Color online) Measured average dissipation and its variance as a function of $\nu$ for $\beta=0$. The solid line corresponds to the theoretical prediction for $d_{\text{av}}$, eq. (\ref{4.18a}), while the dashed line is the gaussian estimation of the dissipation variance parameter, $\Lambda_{\nu}^2$, see eq. (\ref{4.22}). The agreement is excellent in all cases. Notice in particular the scaling with $\nu$ of both observables in the weakly- and strongly-dissipative system limits.
}
\label{average-variance}
\end{figure}

We have tested the above predictions in standard Monte Carlo simulations of the dissipative KMP model described in this section, for the particular case $\beta=0$. Figure \ref{comparisontogaussian} shows the probability density function (pdf) for the dissipated energy, integrated over the whole system and over a long time $\tau$ for many different values of the macroscopic dissipation coefficient $\nu\in[10^{-3},10^3]$. In order to minimize finite-size effects in the measurements, we performed simulations for systems with increasing size as $\nu$ grows, $L\propto \ell_{\nu}^{-1}$, in such a way that the number of lattice sites per unit typical length is constant and large enough so we are within the hydrodynamic regime. Furthermore, the integration time $\tau=\mathcal{O}(1)$ for the continuous, diffusive, timescale over which the hydrodynamic predictions should hold \cite{step}.  Standard Monte Carlo simulations do not allow us to sample the tails of the distribution, but they are useful to study the typical fluctuations around the average we are interested in here (e.g., a regime of 5 standard deviations around the average). Figure \ref{comparisontogaussian} shows that, when plotted against the reduced variable $z\equiv (d-d_{\text{av}})/\chi$, the distribution $P_{\tau}(z)$ follows approximately a normal distribution for typical fluctuations. Moreover, all curves for different $\nu$ collapse in this regime. However, even at this standard simulation level, it becomes apparent that the tails of the distribution (corresponding to moderate dissipation fluctuations) deviate from gaussian behavior, see inset in Fig. \ref{comparisontogaussian}, showing asymmetric tails and breaking the collapse to gaussian behavior observed for small fluctuations. As we will show below, the analysis of the dissipation LDF shows that the large fluctuations statistics is far from gaussian.

In order to further check our theory, we have also compared the measured average dissipation and its variance with the analytical results above, as a function of the macroscopic dissipation coefficient $\nu$, varying in a range which covers 6 orders of magnitude. Again, we see in Fig. \ref{average-variance} that the agreement is excellent in all cases. In particular, the average dissipation grows as $\nu$ (resp. $\nu^{1/2}$) in the weakly (resp. strongly) dissipative system limit, while the variance remains constant for $\nu\ll 1$ but decays as $\nu^{-1/2}$ for $\nu\gg 1$. Remarkably, the gaussian approximation for the variance turns out to be an excellent estimator of the empirical dissipation variance. For $L\tau\gg 1$, large fluctuations of the dissipation are very rare and most of the probability concentrates in a region of width proportional to $(L\tau)^{-1/2}$ around the average value, a regime described by the gaussian approximation.

\subsection{Complete fluctuation spectrum for the integrated dissipation}

We now investigate the whole spectrum of fluctuations (both typical and rare) of the integrated dissipation. Thus, we need to evaluate the LDF $G(d)$ for arbitrary values of $d$, in general not close to its average value $d_{\text{av}}$, both analytically and numerically. Exploring in standard simulations the tails of the dissipation distribution associated to the nontrivial structure of $G(d)$ is an daunting task, since LDFs involve by definition exponentially-unlikely rare events, see eq. (\ref{1.5}). This has been corroborated in Fig. \ref{comparisontogaussian}, where the dissipation distribution has been measured directly but we are unable to gather enough statistics in the tails of the pdf to obtain clear-cut results in the non-gaussian regime. A recent series of works have addressed this issue, developing an efficient method to measure directly LDFs in many particle systems \cite{sim,sim2,sim3}. The method is based on a modification of the dynamics so that the rare events responsible of the large deviation are no longer rare \cite{sim}, and it has been developed for discrete- \cite{sim} and continuous-time Markov dynamics \cite{sim2}. For a recent review, which also discusses Hamiltonian systems, see ref. \cite{sim3}. The method yields the Legendre-Fenchel transform of the dissipation LDF, which is usually defined as  $\mu(s)=\max_d[G(d)+s d]$ \cite{Touchette,foot3}. In particular, if $U_{C' C}$ is the transition rate from configuration $C$ to $C'$ of the associated stochastic process, the modified dynamics is defined as $\tilde{U}_{C' C}(s) = U_{C'C} \exp(s \, d_{C' C})$, where $d_{C'C}$ is the energy dissipated in the elementary transition $C \to C'$. It can be then shown \cite{sim,sim2,sim3,Pablo} that the natural logarithm of the largest eigenvalue of \emph{matrix} $\tilde{U}(s)$ gives $\mu(s)$, which in turn can be Legendre-transformed back to obtain a Monte Carlo estimate of $G(d)$. The method of refs. \cite{sim,sim2,sim3} thus provides a way to measure $\mu(s)$ by evolving a large number $M$ of copies or clones of the system using the modified dynamics $\tilde{U}(s)$. This method is exact in the limit $M\to \infty$, but in practice we are able to simulate a large but finite population of clones, typically $M\in [10^3,10^4]$. This introduces additional finite-size effects related to the population of clones which must be considered with care, see \cite{Pablo3} for further discussion along this line. The numerical results for the LDF in the following sections have been obtained using these advanced Monte Carlo techniques.

\subsubsection{Weakly-dissipative systems, $\nu\ll 1$}
\label{s4b}

We now focus our attention on the analysis of LDF of the integrated dissipation for weakly dissipative systems, in which $\nu\ll 1$. In the general framework developed in section \ref{s1}, we found a scaling property for $G(d)$, as given by eq. (\ref{1.41}),
\begin{equation}\label{4.23}
 \left(\frac{d_{\text{av}}}{\nu}\right)^{\gamma-2}G(d)=- \left[ 1-\frac{d/d_{\text{av}}}{Y_0(\widetilde{\calH})}\right] \widetilde{\mathcal{H}} ,
\end{equation}
where $\gamma=(2+\beta)/(1+\beta)$, $Y_0(\widetilde{\calH})$ is determined by Eq. (\ref{1.34}), and the constant $\widetilde{\calH}$ depends only on the ratio $d/d_{\text{av}}$, as given by eq. (\ref{1.35}).

For the sake of concreteness, let us consider now the simplest case $\beta=0$, corresponding to the dissipative KMP model introduced in \cite{PLyH11a}. Equation (\ref{1.32}) for the rescaled density profile now reads
\begin{equation}\label{4.24}
    Y'(x)^2=8 \widetilde{\calH} Y^2 \left(1-\frac{Y}{Y_0}\right), \quad Y(\pm 1/2)=1,
\end{equation}
which can be explicitly integrated, with the solution
\begin{equation}\label{4.25}
    Y(x,\widetilde{\calH})=Y_0 \sech^2(x\sqrt{2\widetilde{\calH}}), \quad Y_0=\cosh^2 \sqrt{\frac{\widetilde{\calH}}{2}},
\end{equation}
where we have already used that $Y(x)$ must be an even function of $x$. The {rescaled current profile} $\Psi(x)$ introduced in (\ref{1.33}) is
\begin{equation}\label{4.26}
    \Psi(x,\widetilde{\calH})=-\frac{\cosh^2 \sqrt{\frac{\widetilde{\calH}}{2}}}{\sqrt{2\widetilde{\calH}}} \tanh(x\sqrt{2\widetilde{\calH}}).
\end{equation}
The optimal profiles for the density and the current can be now readily written by combining the previous two equations with Eqs. (\ref{1.42})-(\ref{1.43}), yielding
\begin{subequations}\label{4.29}
\begin{equation}\label{4.29a}
  \rho(x)=T Y(x)=T \cosh^2 \sqrt{\frac{\widetilde{\calH}}{2}} \sech^2(x\sqrt{2\widetilde{\calH}}),
\end{equation}
\begin{equation}\label{4.29b}
  j(x)= d_{\text{av}} \Psi(x)=-d_{\text{av}} \frac{\cosh^2 \sqrt{\frac{\widetilde{\calH}}{2}}}{\sqrt{2\widetilde{\calH}}} \tanh(x\sqrt{2\widetilde{\calH}})
\end{equation}
\end{subequations}
in terms of $\widetilde{\calH}=\widetilde{\calH}(d)$. We have taken into account that $y\equiv \rho$ for $\beta=0$. Note that the curves $\rho(x)/T=Y(x,\widetilde{\calH})$ for different values of $\nu$ plotted as a function of $x$ only depend on the relative dissipation $d/d_{\text{av}}$. Now, eq. (\ref{1.35}) implies that
\begin{equation}\label{4.27}
    \frac{d}{d_{\text{av}}}=2 \Psi(-1/2)=
    \frac{\sinh{\sqrt{2\widetilde{\calH}}}}{\sqrt{2\widetilde{\calH}}} , \quad (d_{\text{av}}\sim\nu T),
\end{equation}
which gives the constant $\widetilde{\calH}$ implicitly in terms of $d/d_{\text{av}}$.
Finally, particularizing eq. (\ref{4.23}) for the case $\gamma=2$ we are analyzing (that is, $\beta=0$), we obtain
\begin{equation}\label{4.28}
    G(d)=\sqrt{2\widetilde{\calH}} \tanh{\sqrt{\frac{\widetilde{\calH}}{2}}}-\widetilde{\calH} \, .
\end{equation}
Note that eq. (\ref{4.27}) for $\widetilde{\calH}(d)$ requires some careful analysis. From the general discussion in Section \ref{s1.5b}, we have that $\widetilde{\calH}>0$ for $d>d_{\text{av}}$,  and thus $\sqrt{2\widetilde{\calH}}$ is a real number, while $\widetilde{\calH}<0$ for $d<d_{\text{av}}$ and $\sqrt{2\widetilde{\calH}}$ is imaginary. The latter case poses no problem for $G(d)$, which is always real-valued. In fact, if we write $\sqrt{\widetilde{\calH}}=i\sqrt{|\widetilde{\calH}|}$ we arrive at $G(d)=-\sqrt{2|\widetilde{\calH}|} \tan{\sqrt{\frac{|\widetilde{\calH}|}{2}}}+|\widetilde{\calH}|$ for $\widetilde{\calH}<0$. In the limit as $d\to 0$, we have that $\widetilde{\calH}\to -\pi^2/2$, and thus $G(d)\to-\infty$ as expected on a physical basis.

\begin{figure}
\vspace{-0.5cm}
\centerline{
\includegraphics[width=9cm]{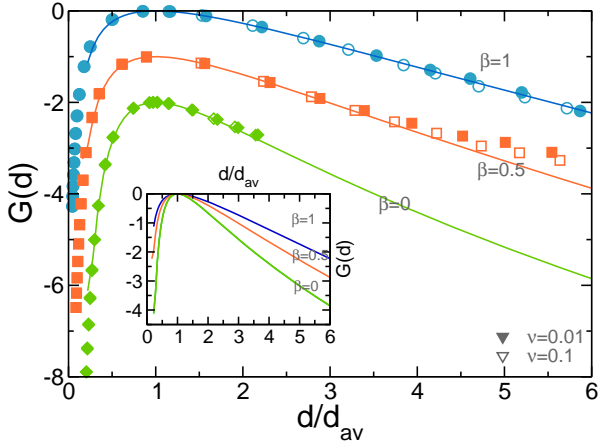}}
\vspace{-0.25cm}
\caption{(Color online) Scaling of the dissipation LDF in the quasi-elastic limit ($\nu\ll1$) for $N=50$, $T=1$ and varying $\beta$ for two different values of $\nu$, namely $\nu=0.01$ (filled symbols) and $\nu=0.1$ (open symbols). The solid lines are the MFT predictions in each case. Curves have been shifted vertically for convenience, $G(d_{\text{av}})=0$, $\forall \nu,\beta$. {For the case $\beta=0$, the simulation curves are plotted for $d<d_I$, with $d_I$ being the inflection point at which $G(d)$ changes convexity in the limit $\nu\ll 1$ (see the text and also Fig. \ref{muldfsmallnu}).} Inset: Comparison of the theoretical $G(d)$ for different $\beta$, where it is clear that increasing $\beta$ favors larger dissipation fluctuations.}
\label{ldfsmallnu}
\end{figure}

\begin{figure}
\vspace{-0.5cm}
\centerline{
\includegraphics[width=9cm]{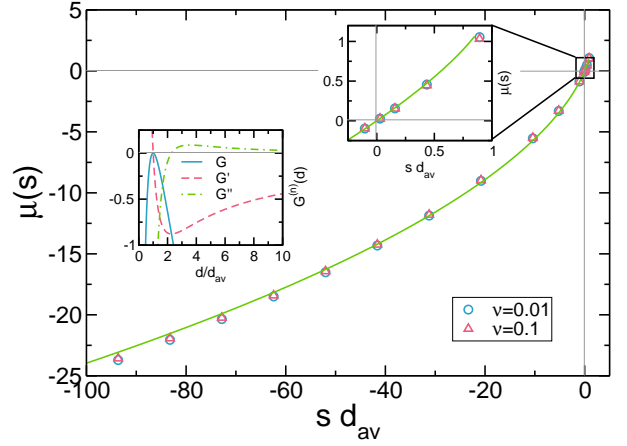}}
\vspace{-0.25cm}
\caption{(Color online) Scaling plot of the Legendre transform of the dissipation LDF, $\mu(s)=\max_d[G(d)+sd]$, in the quasi-elastic limit $\nu\ll1$ for $N=50$, $T=1$, $\beta=0$ and two different values of $\nu$, namely $\nu=0.01$ (circles) and $\nu=0.1$ (triangles). The solid line is the MFT prediction, see eq. (\ref{leg2b}). Notice that $\mu(s)$ is defined up to a threshold value $s_I=0.878458/d_{\text{av}}$, beyond which the Legendre-Fenchel transform diverges. The top-right inset shows a zoom around the threshold $s_I$. This is related to the existence of an inflection point in $G(d)$ for $d_I=2.27672 d_{\text{av}}$, i.e. a point at which $G''(d_I)=0$, see middle-left inset, beyond which the dissipation LDF is non-convex, see discussion in main text.
}
\label{muldfsmallnu}
\end{figure}

Equation (\ref{4.28}) gives a simple scaling form for $G(d)$, independent of $\nu$, for the linear ($\beta=0$) dissipative KMP model in the low-dissipation limit $\nu\ll 1$ \cite{PLyH11a}. As anticipated by (\ref{4.23}), the curve of $G(d)$ vs the relative dissipation $d/d_{\text{av}}$ is independent of $\nu$ in this quasi-elastic regime. This scaling is fully confirmed in Fig. \ref{ldfsmallnu}, in which we plot $G(d)$ for different, small values of $\nu\in[10^{-2},10^{-1}]$ measured in simulations of the dissipative KMP model using the advanced Monte Carlo technique described at the beginning of this subsection. In particular, the agreement between theory and simulations is excellent in the broad fluctuation regime that we could measure (see below). The dissipation LDF is highly skewed with a fast decrease for fluctuations $d<d_{\text{av}}$ and no negative branch, so fluctuation theorem-type relations linking the probabilities of a given integrated dissipation $d$ and the inverse event $-d$ do not hold \cite{GC,LS}. This was of course expected from the lack of microreversibility, a basic tenet for the fluctuation theorem to apply \cite{Puglisi}. The limit $\widetilde{\calH}\gg 1$ corresponds to large dissipation fluctuations, where $G(d) \approx -\frac{1}{2}[\ln(d/d_{\text{av}})]^2$, that is, a very slow decay which shows that such large fluctuations are far more probable than expected within gaussian statistics ($\sim -\frac{3}{2}(d/d_{\text{av}})^2$). In fact, such slow decay implies the presence of an inflection point in $G(d)$: { there is a value $d_I$ such that $G''(d_I)=0$. The convexity of $G(d)$ changes at $d=d_I$, $G''(d)<0$ for $d<d_I$ while $G''(d)>0$ for $d>d_I$. Specifically, Eqs. (\ref{4.27}) and (\ref{4.28}) imply that}  $d_I/d_{\text{av}}=2.27672$ (see middle-left inset in Fig. \ref{muldfsmallnu}). {The complete measurement of} non-convex LDFs in computer simulations is a challenge which remains unsolved. The reason is that the advanced Monte Carlo method described above to directly measure large-deviation functions in simulations is based on the Legendre-Fenchel transform for the LDF of interest, which is not well-behaved in regimes where the LDF is non-convex \cite{Touchette}.

To better understand this issue, recall that the Legendre-Fenchel transform of the dissipation LDF can be written as
\be
\mu(s)=\max_d[G(d)+s d] = G[d^*(s)]+ s \, d^*(s) \, ,
\label{muldf}
\ee
where $d^*(s)$ is solution of the equation
\be
\frac{\partial G(d)}{\partial d}=-s \, .
\label{muldfeq}
\ee
{Note that, mathematically, $\mu(s)$ is the Legendre-Fenchel transform of $-G(d)$, because the Legendre-Fenchel transform is defined for convex functions \cite{Touchette}. The} partial derivative of $G$ with respect to $d$ is related to the first integral {of Hamilton equations} $\Pi_{\psi 0}$, see Eq. (\ref{var13}), which in turn can be obtained from eqs. (\ref{1.44}) and (\ref{4.25}), yielding
\begin{equation}
\Pi_{\psi 0}=\nu \frac{\partial G}{\partial d}=-\frac{\widetilde{\calH}}{T} \sech^2 \sqrt{\frac{\widetilde{\calH}}{2}}.
\label{4.30}
\end{equation}
Equivalently,
\begin{equation}
s=-\frac{\partial G}{\partial d}=\frac{\widetilde{\calH}}{\nu T} \sech^2 \sqrt{\frac{\widetilde{\calH}}{2}} \, .
\label{4.31}
\end{equation}
In this way, making use of Eqs. (\ref{4.27}), (\ref{4.28}) and (\ref{4.31}), the Legendre transform of the dissipation LDF can be written as
\be
\mu(s)=2\sqrt{2\widetilde{H}}  \tanh\sqrt{\frac{\widetilde{H}}{2}} - \widetilde{\calH} \, ,
\label{leg2b}
\ee
in terms of $\widetilde{\calH}$, which is obtained implicitly {as a function of $s$} from Eq. (\ref{4.31}). Note that the scaling of $G(d)$ with $d/d_\text{av}$, see Eq. (\ref{4.23}), implies a similar collapse for $\mu(s)$ when plotted as a function of $s \, d_\text{av}$. Eq. (\ref{muldfeq}) has a single solution $d^*(s)$ for $s<0$ and hence poses no problem. On the other hand, due to the existence of an inflection point, $G'(d)$ exhibits a minimum at $d_I$, increasingly smoothly afterward to reach asymptoticaly zero in the limit $d\to \infty$, see middle-left inset in Fig. \ref{muldfsmallnu}. Therefore, for $s>0$ there exist two solutions $d_1^*(s)\le d_I\le d_2^*(s)$ for eq. (\ref{muldfeq}), but only the first one maximizes eq. (\ref{muldf}). This means that we cannot obtain $G(d)$ by inverse Legendre-transforming $\mu(s)$ for dissipations above the inflection point $d_I=2.27672 \, d_{\text{av}}$. In fact, $\mu(s)$ is defined up to a critical $s_I$, such that $s_I=0.87845/d_{\text{av}}$ (the slope of $-G(d)$ at the inflection point), beyond which $\mu(s)$ diverges. This can be seen by noticing the main properties of $\mu(s)$, namely
\begin{equation}\label{leg3}
  \frac{\partial\mu}{\partial s}=d \, , \quad \frac{\partial^2\mu}{\partial s^2}=-\left[ \frac{\partial^2 G}{\partial d^2}\right]^{-1}.
\end{equation}
Therefore, $\mu$ has a singularity at the value of the slope $s_I$ corresponding to the inflection point $d_I$, where $\partial^2\mu/\partial s^2$ diverges. The transition to non-convex behavior thus implies that we can only measure the statistics of rare dissipation fluctuations up to $d_I$ using the cloning algorithm \cite{sim,sim2,sim3}. Fig. \ref{muldfsmallnu} shows a comparison between the measured $\mu(s)$ for two different values of $\nu\ll 1$ and the theoretical expectation, up to the critical $s_I$. The agreement is excellent in all cases, and the collapse of $\mu(s)$ when plotted against $s \, d_\text{av}$ is confirmed. The challenge remains to devise computational techniques capable of exploring rare-event statistics even in regimes where the associated LDF is non-convex.

\begin{figure}
\vspace{-0.5cm}
\centerline{
\includegraphics[width=9cm]{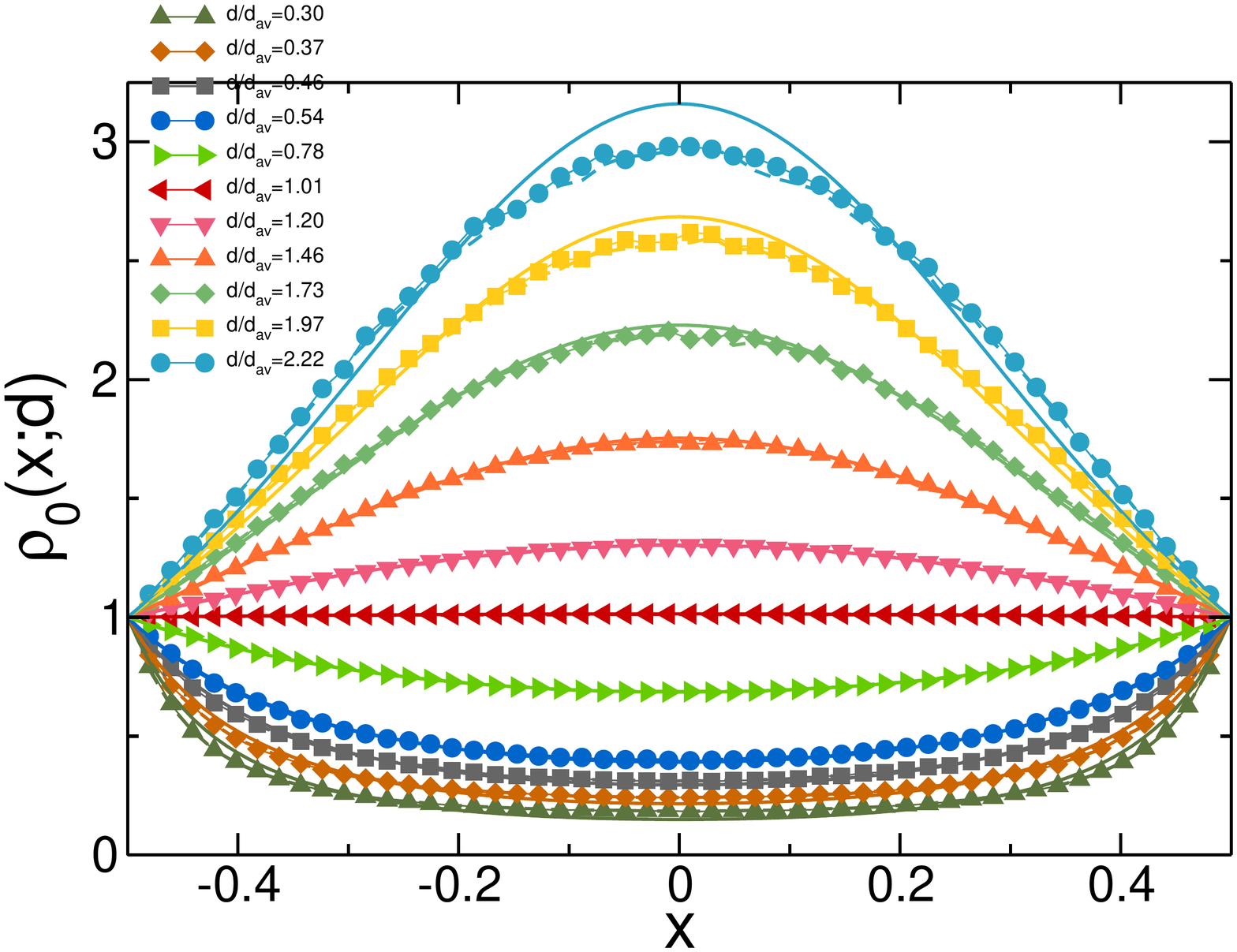}}
\centerline{
\includegraphics[width=7.5cm]{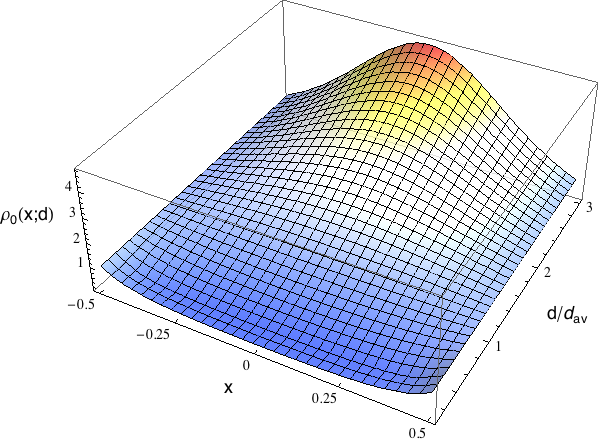}}
\vspace{-0.25cm}
\caption{(Color online) Top: Optimal energy profiles for varying $d/d_{\text{av}}$ and $\beta=0$, measured for $\nu=10^{-3}$ (symbols) and $\nu=10^{-2}$ (dashed lines), and MFT predictions (solid lines). Agreement is very good in all cases. Bottom: MFT prediction for the optimal density profiles for varying $d/d_{\text{av}}$.
}
\label{profsmallnu}
\end{figure}

We may solve in a similar way the MFT for the integrated dissipation for arbitrary values of the exponent $\beta$, though mathematical expressions are far more convoluted that in the illustrative case $\beta=0$ described above. Fig. \ref{ldfsmallnu} also shows the dissipation LDF for other exponents $\beta>0$, as well as the results of numerical experiments in these cases. Qualitatively, the results are equivalent to those discussed above, with a $\nu$-independent scaling form of the LDF in the $\nu\ll 1$ limit which goes rapidly to zero as $d\to 0$ and has a relatively fat tail for $d\gg d_{\text{av}}$. {This tail changes} convexity (based on a numerical analysis) at a large dissipation $d_I$ which increases with {(a) $\nu$ for fixed $\beta$ (b) $\beta$ for fixed $\nu$. For $\beta$=0, we have $d_I/d_{\text{av}}\simeq 2.8$ for $\nu=1$, while $d_I/d_{\text{av}} > 6$ for $\nu=10$. On the other hand, for $\beta=1$, $G''(d)<0$ in the considered region and no inflection point therein.  For the intermediate value, $\beta=0.5$, the positive values of $G''(d)$ are so small that we have chosen not to eliminate the points behind the numerical inflection point, $d_I/d_{\text{av}}\simeq 3.2$ for $\nu\ll 1$ and $d_I/d_{\text{av}}\simeq 5.1$ for $\nu=1$, although they roughly coincide with the values at which the theoretical and the simulation curves begin to separate.} Furthermore, comparison with numerical results is excellent in all cases. Interestingly, see inset in Fig. \ref{ldfsmallnu}, increasing $\beta$ results in a broader dissipation LDF, meaning that large dissipation fluctuations are enhanced as $\beta$ grows away from the linear case $\beta=0$.

\begin{figure}
\centerline{
\includegraphics[width=9cm]{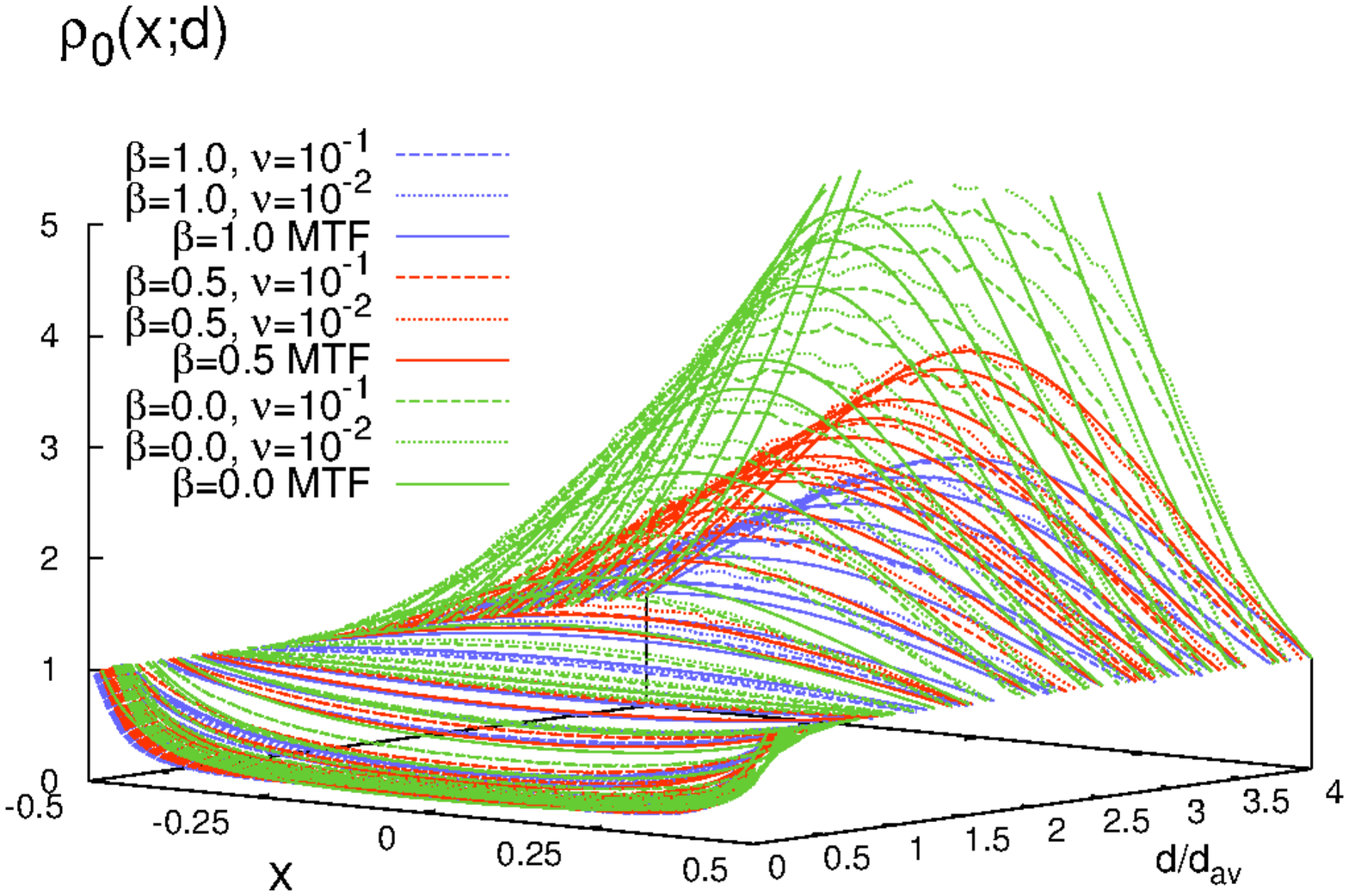}}
\vspace{-0.5cm}
\caption{(Color online) Collapse of the optimal energy profiles measured for $N=50$ and $T=1$ as a function of the relative dissipation $d/d_{\text{av}}$ for different values of $\nu\ll 1$, namely $\nu=10^{-2}$ (dotted lines) and $\nu=10^{-1}$ (dashed lines), for $\beta=0$ (top, green), $\beta=0.5$ (middle, red) and $\beta=1$ (bottom, blue). The larger $\beta$, the less pronounced the central overshoot is for $d>d_{\text{av}}$. Solid lines correspond to MFT predictions.
}
\label{profsmallnubeta}
\end{figure}

We have also measured the typical energy profile associated to a given dissipation fluctuation for the case $\beta=0$, see top panel in Fig. \ref{profsmallnu}, finding also very good agreement with the macroscopic fluctuating theory developed in this paper. Remarkably, optimal profiles for varying $\nu\ll 1$ also collapse for constant $d/d_{\text{av}}$ (all the simulations have been done with the same value of the energy density at the boundaries $T=1$), as predicted by eq. (\ref{4.29a}). Furthermore, profiles exhibit the $x\leftrightarrow -x$ symmetry conjectured in section \ref{s1.2b} in all cases, with a single extremum which can be minimum or maximum depending on the value of the relative dissipation $d/d_{\text{av}}$, a property which was deduced from the general formalism in section \ref{s1.5b}. Interestingly, profiles associated to dissipation fluctuations above the average exhibit an energy overshoot in the bulk. This observation suggests that the mechanism responsible for large dissipation fluctuations consists in a continued over-injection of energy from the boundary bath, which is transported to and stored in the bulk before being dissipated. The same qualitative observations and good agreement between theory and simulations is observed for other values of the exponent $\beta>0$, see Fig. \ref{profsmallnubeta}. Notice in particular the nice collapse of optimal profiles for different values of $\nu\ll 1$ but equal relative dissipation. An interesting observation is that optimal density profiles are less pronounced the larger de nonlinearity exponent $\beta$ is, see Fig. \ref{profsmallnubeta}. This gives a plausible explanation of the widening of $G(d)$ as $\beta$ increases: for the same value of $d/d_{\text{av}}$ and increasing $\beta$, the associated optimal profile is closer to the hydrodynamic solution the larger $\beta$ is, and hence this fluctuation cost decreases, having a larger associated probability.

\begin{figure}
\vspace{-0.5cm}
\centerline{
\includegraphics[width=9cm]{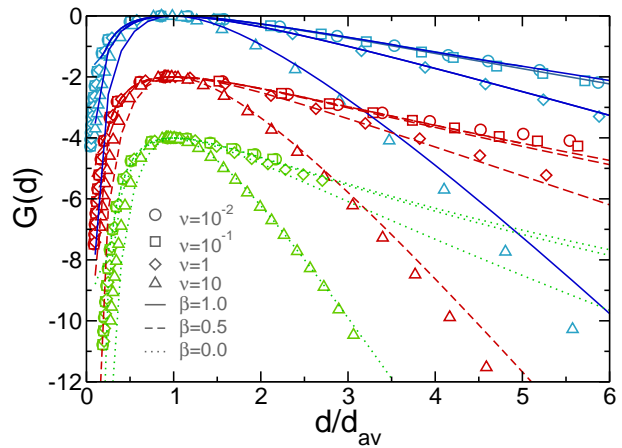}}
\vspace{-0.25cm}
\caption{(Color online) Dissipation LDF for $N=50$, $T=1$ and varying $\beta=0, \, 0.5,\, 1.0$ and $\nu\in[10^{-2},10]$. Curves for $\beta=0.5$ and $0$ have been shifted vertically for convenience (recall that $G(d_{\text{av}})=0 \,\, \forall \nu,\beta$), so that $\beta=1$, $0.5$ and $0$ correspond to top (blue), medium (red) and bottom (green). The MFT predictions are plotted with lines: solid for $\beta=1$, dashed for $\beta=0.5$, and dotted for $\beta=0$. As in Fig. \ref{ldfsmallnu}, for a fixed $\nu$ increasing $\beta$ results in larger dissipation fluctuations.}
\label{ldfall}
\end{figure}

\subsubsection{Arbitrary dissipation coefficient $\nu$}

For arbitrary values of $\nu \gtrsim 1$ no general scaling function can be derived in principle for $G(d)$. For each particular case, the whole variational problem, eqs. (\ref{4.14})-(\ref{4.15e}), must be solved, which is often analytically intractable. In order to further advance, we resort now to a numerical evaluation of the optimal profiles, which are used in turn to compute the dissipation LDF. Fig. \ref{ldfall} shows the theoretical predictions for $G(d)$ for increasing, non-perturbative values of $\nu$, together with numerical results from simulations, for different values of $\beta$. As for the weakly-dissipative system limit previously discussed, the agreement between theory and measurements in Fig. \ref{ldfall} is quite good. We attribute the observed differences between theory and simulation to finite size effects in the latter, which are more apparent for large $\nu$ as compared to the weakly-dissipative system limit $\nu\ll 1$, compare with Fig. \ref{ldfsmallnu}, {see also \cite{PLyH12a}}. Such strong finite-size effects are expected since the natural length scale associated to a given $\nu$ is $\ell_{\nu}=\sqrt{\hat{D}/\nu}$. {As follows from Eq. (\ref{4.12}) and the associated discussion}, $\ell_\nu$ decreases as $\nu$ grows so larger system sizes are needed to observe convergence to the macroscopic limit. In addition, finite-size effects related to the number of clones $M$ used for the sampling become an issue in this limit \cite{Pablo3,PabloSSB}.

\begin{figure}
\centerline{
\includegraphics[width=8cm]{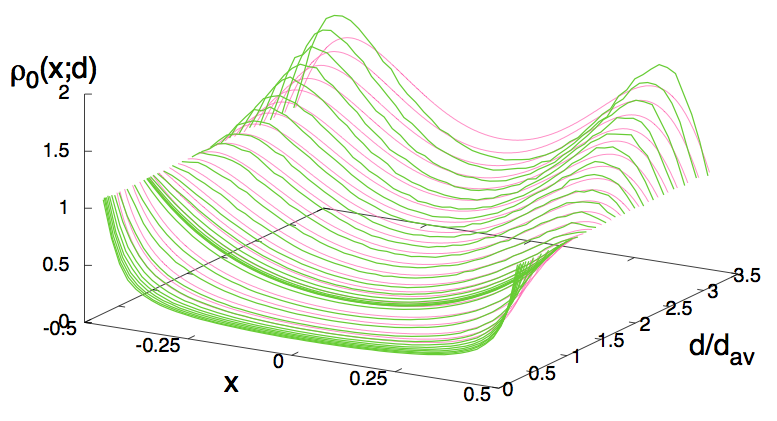}}
\centerline{
\includegraphics[width=8cm]{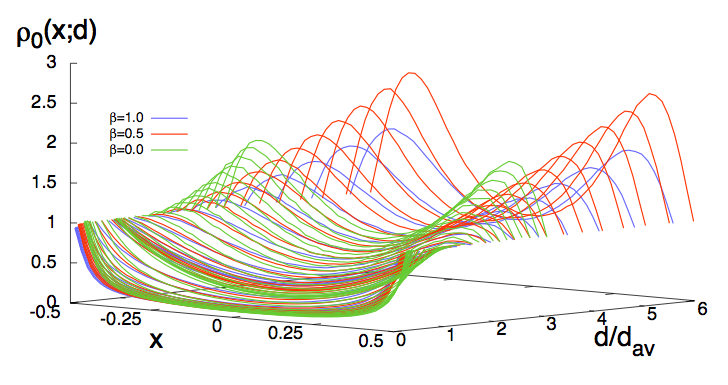}}
\vspace{-0.25cm}
\caption{(Color online) Top: Optimal energy profiles as a function of the relative dissipation measured for $\nu=10$, $N=50$ and $T=1$ for the particular case $\beta=0$. Thick (green) lines correspond to measurements while thin (pink) lines are MFT predictions. Bottom: Measured optimal energy profiles for $\nu=10$, $N=50$ and $T=1$, and varying values of the nonlinearity exponent $\beta$. For a given relative dissipation, energy localization around thermal baths decreases as $\beta$ increases.
}
\label{profcompnu10}
\end{figure}

In any case, the sharpening of $G(d)$ as $\nu$ increases for any $\beta$ shows that large dissipation fluctuations are strongly suppressed in this regime, as was argued for $\nu\gg 1$ on quite general grounds in Sec. \ref{s1.5c}). In this strongly-dissipative system limit $\nu\gg 1$ the scale $\ell_{\nu}\to 0$, and the system decouples effectively into two independent boundary shells. Thus, the energy is concentrated around the boundary baths, a picture which agrees again with the analysis of Sec. \ref{s1.5c}. This behaviour is evidenced by the optimal energy profiles for a given $d$ measured for $\nu=10$, see Fig. \ref{profcompnu10}, in contrast to the behavior observed for $\nu\ll 1$, see Figs. \ref{profsmallnu}-\ref{profsmallnubeta}. The agreement of the observed profiles with MFT predictions is rather good, taking into account the non-negligible finite-size effects affecting these measurements. Bottom panel in Fig. \ref{profcompnu10} shows the measured energy profiles as a function of the relative dissipation and for different values of the nonlinearity exponent $\beta$. From this figure, it is clear that for a given relative dissipation, energy localization around thermal baths decreases as $\beta$ increases. This suggests again that, as in the $\nu\ll 1$ limit, the probability of a fixed relative dissipation fluctuation $d/d_{\text{av}}$, increases as $\beta$ grows, giving rise to a broadening of $G(d)$ with $\beta$.

\section{Summary and conclusions}
\label{s5}

In this paper we have developed a general theoretical framework for calculating the probability of large deviations for the dissipated energy in a general class of nonlinear driven diffusive systems with dissipation. Our starting point is a mesoscopic fluctuating hydrodynamic theory for the energy density in terms if a few slow hydrodynamic fields, that is, a fluctuating reaction-diffusion equation with a drift term compatible with Fourier's law and a sink term which can be written in terms of the local energy density. The validity of this hydrodynamic description can be demonstrated for a large family of stochastic microscopic models \cite{PLyH12a}, but it is expected to describe the coarse-grained physics of many real systems sharing the same main ingredients, namely: (i) nolinear diffusive dynamics, (ii) bulk dissipation, and (iii) boundary driving. From this fluctuating hydrodynamic description, and using a standard path integral formulation of the problem, we can write the probability of a path in mesoscopic phase space, that is, the space spanned by the slow hydrodynamic fields. Interestingly, the action associated to this path, from which large-deviation functions for macroscopic observables can be derived, has the same simple form as in non-dissipative systems. This is a consequence of the quasi-elasticity of microscopic dynamics, required in order to have a nontrivial competition between diffusion and dissipation at the mesoscale \cite{PLyH12a}.

We use the derived action functional to investigate the large deviation function of the dissipated energy. The energy dissipated in a non-conserving diffusive system is, together with the energy current, the relevant macroscopic observable characterizing nonequilibrium behavior. A simple and powerful additivity conjecture simplifies the resulting variational problem for the dissipation LDF, from which we arrive at Euler-Lagrange equations for the optimal density and current fields that sustain an arbitrary dissipation fluctuation. A Hamiltonian reformulation of this variational problem greatly simplifies the calculations, allowing us to analyze the general theory in certain interesting limits. A perturbative solution thereof shows that the probability distribution of small (that is, typical) fluctuations of the dissipated energy is always gaussian, as expected from the central limit theorem. Moreover, a general expression for the variance of the distribution in the gaussian approximation has been derived which compares nicely with numerical results. On the other hand, strong separation from the gaussian behavior is expected for large dissipation fluctuations, with a distribution which shows no negative branch, thus violating the Gallavotti-Cohen fluctuation theorem as expected from the irreversibility of the dynamics. Furthermore, the dissipation LDF exhibits simple and general scaling forms in the weakly- and strongly-dissipative system limits, which can be analyzed in general without knowing the explicit solution of the canonical equations.

We apply our results to a general class of diffusive lattice models for which dissipation, nonlinear diffusion and driving are the key ingredients. The theoretical predictions, which can be explicitely worked out in certain cases, are compared to extensive numerical simulations of the microscopic models (which cover both typical fluctuations and rare events), and excellent agreement is found in all cases. In particular, the simple scaling for the dissipation large-deviation function in the weakly-dissipative system limit is fully confirmed for different values of the nonlinearity exponent $\beta$, exhibiting non-convex behavior for large enough fluctuations. Interestingly, in this limit $\nu\ll 1$ energy profiles associated to large dissipation fluctuations exhibit an overshoot in the bulk resulting from an excess energy injection from boundary baths. On the other hand, in the strongly-dissipative system limit $\nu\gg 1$ the typical lengthscale goes to zero and the system decouples into two almost-independent boundary shells, giving rise to a different scaling form for the LDF and a strong suppression of the dissipation fluctuations in this regime.

Recently, a similar hydrodynamic theory has been developed to study large fluctuations in a particular class of driven dissipative media \cite{Bodineau}, but its predictions do not compare well with numerical results of the dissipative lattice models here studied. The reason is that Ref. \cite{Bodineau} studies systems with two competing dynamics, one conservative and another nonconservative, thus resulting in independent fluctuations for the density and dissipation fields. In our theory, as is the case in many driven dissipative systems, dissipation is linked to the collision process, and hence dissipation fluctuations are enslaved to density profile deviations. In fact, both theories coincide in the limit where the optimal dissipation profile is given in terms of the optimal density field.

In summary, our results show that a suitable generalization of macroscopic fluctuation theory \cite{Bertini} is capable of describing in detail the fluctuating behavior of general nonlinear driven dissipative media. In this scheme, the dissipation LDF follows from a variational problem whose solution also gives the optimal profiles that the system has to sustain to achieve the considered fluctuation. The proposed framework is very general, as MFT is based only on (a) the knowledge of the conservation laws governing a system, which allow to write down the balance equations for the fluctuating fields, and (b) a few transport coefficients appearing in these fluctuating balance equations. This opens the door to further general results in the nonequilibrium statistical physics of dissipative media. In particular, it would be interesting to explore the existence of phase transitions and spontaneous symmetry breaking at the fluctuating level therein, in a way similar to the phenomenon reported in conservative systems \cite{BD05,PabloSSB}. Moreover, as the relevant magnitudes characterizing nonequilibrium behavior in dissipative systems are both the dissipated energy and the current, it would be worth analyzing the joint fluctuations of these two observables within the MFT approach.

\acknowledgments
We acknowledge financial support from Spanish Ministerio de Ciencia e Innovaci\'on projects FIS2011-24460 and FIS2009-08451, EU-FEDER funds, and Junta de Andaluc\'{\i}a projects P07-FQM02725 and P09-FQM4682.

\appendix*

\section{Variational problem with a Lagrangian including second-order derivatives}

Let us analyze a variational problem in which the ``action'' is defined as the integral of a ``Lagrangian'' with second order derivatives, that is
\begin{equation}\label{ap1}
  \calS[j]=\int_{x_1}^{x_2} dx \, \calL(j,j',j'').
\end{equation}
The action $\calS[j]$ is a functional of the profile $j(x)$ in the fixed interval $x_1\leq x \leq x_2$. The variational problem arises when one looks for the ``optimal'' profile $j(x)$ for which the functional $\calS[j]$ is a extremum. For the sake of concreteness, let us consider a problem similar to the one analyzed in this paper: we are interested in calculating $G$ defined as
\begin{equation}\label{ap10}
  G= -\min_{j(x)} S[j].
\end{equation}
Then, we consider the variation $\delta\calS$ of the functional when a given profile $j(x)$ is slightly changed to $j(x)+\delta j(x)$,
\begin{equation}\label{ap2}
  \delta\calS=\int_{x_1}^{x_2} dx\, \left(\frac{\partial\calL}{\partial j}\delta j+\frac{\partial\calL}{\partial j'}\delta j'+\frac{\partial\calL}{\partial j''}\delta j'' \right).
\end{equation}
Now, we take into account that
\begin{equation}\label{ap3}
  \delta j'=\frac{d}{dx} \delta j, \quad \delta j''=\frac{d^2}{dx^2} \delta j
\end{equation}
in order to integrate by parts (i) once the term proportional to $\delta j'$ (ii) twice the term proportional to $\delta j''$. We arrive thus at
\begin{widetext}
\begin{equation}\label{ap4}
  \delta\calS= \left\{ \left[\frac{\partial\calL}{\partial j'}-\frac{d}{dx}\left(\frac{\partial\calL}{\partial j''}\right)\right] \delta j+ \frac{\partial\calL}{\partial j''} \delta j' \right\}_{x_1}^{x_2} +\int_{x_1}^{x_2}dx\, \left[\frac{\partial\calL}{\partial j}-\frac{d}{dx}\left(\frac{\partial\calL}{\partial j'}\right)+\frac{d^2}{dx^2}\left(\frac{\partial\calL}{\partial j''}\right)\right],
\end{equation}
where $[f]_{x_1}^{x_2}=f(x_2)-f(x_1)$. By analogy with the case of the usual Lagrangian with only first-order derivatives, we introduce the generalized momenta as
\begin{equation}\label{ap5}
  p_j=\frac{\partial\calL}{\partial j'}-\frac{d}{dx}\left(\frac{\partial\calL}{\partial j''}\right), \quad p_{j'}=\frac{\partial\calL}{\partial j''}.
\end{equation}
In this way, the boundary term has the usual form and Eq. (\ref{ap4}) can be rewritten as
\begin{equation}\label{ap6}
  \delta\calS= \left[ p_j \delta j+ p_{j'} \delta j' \right]_{x_1}^{x_2} +\int_{x_1}^{x_2}dx\, \left[\frac{\partial\calL}{\partial j}-\frac{d}{dx}\left(\frac{\partial\calL}{\partial j'}\right)+\frac{d^2}{dx^2}\left(\frac{\partial\calL}{\partial j''}\right)\right] \delta j.
\end{equation}
\end{widetext}

The extremum condition is $\delta S=0$. If the values of $j$ and $j'$ are prescribed at the boundaries, both $\delta j$ and $\delta j'$ vanish at $x_{1,2}$ and the boundary term vanishes. Then, as $\delta j$ is arbitrary for $x_1<x<x_2$, the ``optimal'' profile solution of the variational problem verifies the Euler-Lagrange equation
\begin{equation}\label{ap8}
  \frac{d^2}{dx^2}\left(\frac{\partial \cal L}{\partial j''}\right)-\frac{d}{dx}\left( \frac{\partial\cal  L}{\partial j'}\right)+\frac{\partial \cal L}{\partial j} =0,
\end{equation}
which is a fourth-order differential equation. Interestingly, Eq. (\ref{ap8}) can be written as $dp_j/dx=\partial\calL/\partial j$ that is formally identical to the usual Euler-Lagrange equation for Lagrangians with only first-order derivatives. The boundary conditions for the Euler-Lagrange equation are the prescribed values of $j$ and $j'$ at the boundaries; four conditions for the fourth-order differential equation. However, in physical problems there are sometimes less prescribed quantities at the boundaries than necessary. In that case, as pointed out by Lanczos \cite{La49}, the extremum condition $\delta\calS=0$ provides the ``missing'' boundary conditions. For instance, if we only have fixed values of $j'$ at the boundaries (as in the LDF problem we have dealt with in the main text), $\delta j'(x_1)=\delta j'(x_2)=0$ but $\delta j(x_1)$ and $\delta j(x_2)$ are free parameters. Equation (\ref{ap6}) still implies the Euler-Lagrange equation but also that
\begin{equation}\label{ap9}
  p_j(x_1)=p_j(x_2)=0.
\end{equation}
The generalized momentum conjugate of the variable that is not fixed at the boundary must vanish: the solution of the variational problem verifies then the Euler-Lagrange equation (\ref{ap8}) with the prescribed values of $j'$ at the boundaries and the ``extra'' conditions provided by Eq. (\ref{ap9}). In this way, we obtain the four conditions needed to determine completely the solution of the Euler-Lagrange equation.

The function $G$ defined in Eq. (\ref{ap10}) depends on the values of $j$ and $j'$  at the boundaries. Making use of Eq. (\ref{ap6}), and taking into account that the optimal profile $j(x)$ verifies the Euler-Lagrange equation, we get
\begin{equation}\label{ap11}
  \delta G=-p_{j,2} \delta j_2- p_{j',2} \delta j'_2+p_{j,1} \delta j_1+ p_{j',1} \delta j'_1.
\end{equation}
We have introduced the notation $p_{j,i}=p_j(x_i)$, $\delta j_i=\delta j(x_i)$, $i=1,2$ and so on. Equation (\ref{ap11}) implies that
\begin{equation}\label{ap12}
  p_{j,2}=-\frac{\partial G}{\partial j_2}, \; p_{j',2}=-\frac{\partial G}{\partial j'_2}, \quad p_{j,1}=\frac{\partial G}{\partial j_1}, \; p_{j',1}=\frac{\partial G}{\partial j'_1}.
\end{equation}
Equation (\ref{var12}) of the main paper is the particularization of this result for the case (i) $x_2=-x_1=1/2$, (ii) solutions of the Euler-Lagrange equation with well-defined parity, in which $p_{j,2}=p_{j,1}$, $p_{j',2}=-p_{j',1}$, and (iii) the boundary conditions of Eq. (\ref{1.11}).


\begin{thebibliography}{99}
\bibitem{foot1} A mathematically similar situation was found in Ref. \cite{KyM12} for the fluctuations of the current in conservative systems, where a ``standard'' variational problem with a Lagrangian with only first-order derivatives of the current $j$ arose. Thus, in the canonical description, there was only one canonical momentum $p_j$, and the particular solution of the canonical equations with $p_j=0$ also corresponded to the average behaviour.
\bibitem{foot2} Multiple maxima (minima), if present, must be separated by intermediate minima (maxima), but this is not possible as $y(x)$ takes the same value $y_0$ at all these extrema. The possibility of an infinite (continuum) set of extrema is ruled out because of the smoothness of $Q(y)$ for the typical systems of interest.
\bibitem{gamma} $\Gamma(z)$ is the gamma function, $\Gamma(z)=\int_0^\infty dt \,t^{z-1} e^{-t}$, with the property $\Gamma(z+1)=z\Gamma(z)$
\bibitem{foot3} The parameter $s$ appearing in the Legendre transform of G(d) is conjugated to the dissipation via the equation $s=-\partial G/\partial d$, and turns out to be just minus the Lagrange multiplier introduced in Sec. \ref{s1.2a}, $s=-\lambda$.
\end{thebibliography}
\end{document}